\documentstyle[amstex]{article}
\catcode`\@=11
\def\numberbysection{\@addtoreset{equation}{section}
        \def\theequation{\thesection.\arabic{equation}}}
\numberbysection

\begin{document}

\title{{\huge Bethe Ansatz solution for quantum spin-}${\huge 1}$ {\huge %
chains with boundary terms}{\bf {\huge \ }}}
\author{{\bf E.C. Fireman, A. Lima-Santos and W.Utiel} \\
%EndAName
{\normalsize Universidade Federal de S\~{a}o Carlos, Departmento de F\'{\i}%
sica}\\
{\normalsize Caixa Postal }${\normalsize 676}${\normalsize , CEP }$%
{\normalsize 13569-905}${\normalsize , S\~{a}o Carlos, Brasil }}
\maketitle

\begin{abstract}
The procedure for obtaining integrable open spin chain Hamiltonians via
reflection matrices is explicitly carried out for some three-state vertex
models. We have considered the $19$-vertex models of Zamolodchikov-Fateev
and Izergin-Korepin, and the $Z_{2}$-graded $19$-vertex models with $sl(2|1)$
and $osp(1|2)$ invariances. In each case the eigenspectrum is determined by
application of the coordinate Bethe Ansatz.
\end{abstract}

\vskip1truecm PACS: 05.20.-y, 05.50.+q, 04.20.Jb.

\noindent Keywords: Coordinate {\small BA}, Reflection equation, Open spin
chains.

\section{Introduction}

One-dimensional quantum spin chain Hamiltonians and classical statistical
systems in two spatial dimensions on a lattice (vertex models), share a
common mathematical structure responsible by our understanding of these
integrable models \cite{Baxter, KIB, ABR}. If the Boltzmann weights
underlying the vertex models are obtained from solutions of the Yang-Baxter (%
{\small YB}) equation the commutativity of the associated transfer matrices
immediately follow, leading to their integrability.

The Bethe Ansatz ({\small BA}) is a powerful method in the analysis of
integrable quantum models. There are several versions: coordinate {\small BA}
\cite{Bethe}, algebraic{\small \ BA} \cite{FT}, analytical {\small BA} \cite%
{VR}, etc. The simplest version is the coordinate {\small BA.}\ In this
framework one can obtain the eigenfunctions and the spectrum of the
Hamiltonian from its eigenvalue problem. It is really simple and clear for
the two-state models like the six-vertex models but becomes tricky for
models with a higher number of states.

The algebraic {\small BA}, also known as Quantum Inverse Scattering method,
is an elegant and important generalization of the coordinate {\small BA}. It
is based on the idea of constructing eigenfunctions of the Hamiltonian via
creation and annihilation operators acting on a reference state. Here one
uses the fact that the {\small YB} equation can be recast in the form of
commutation relations for the matrix elements of the monodromy matrix which
play the role of creation and annihilation operators. From this monodromy
matrix we get the transfer matrix which commutes with the Hamiltonian.

Imposing appropriate boundary conditions the {\small BA} method leads to a
system of equations, the Bethe equations, which are useful in the
thermodynamic limit. The energy of the ground state and its excitations,
velocity of sound, etc., may be calculated in this limit. Moreover, in
recent years we witnessed another very fruitful connection between the 
{\small BA} method and conformal field theory. \ Using the algebraic {\small %
BA}, Korepin \cite{KO} found various representations of correlators in
integrable models and more recently Babujian and Flume \cite{BF} developed a
method from the Algebraic {\small BA} which reveals a link to the Gaudin
model, rendering solutions of the Knizhnik-Zamolodchikov equations for the $%
SU(2)$ Wess-Zumino-Novikov-Witten conformal theory in the quasiclassical
limit.

Integrable quantum systems containing Fermi fields have been attracting
increasing interest due to their potential applications in condensed matter
physics. The prototypical examples of such systems are the supersymmetric
generalizations of the Hubbard and $t$-$J$ models \cite{EK0}. They lead to a
generalization of the {\small YB} equation associated with the introduction
of the a $Z_{2}$ grading \cite{KS1} which leads to appearance of additional
signs in the {\small YB} equation.

When considering systems on a finite interval with independent boundary
conditions at each end, we have to introduce reflection matrices to describe
such boundary conditions. Integrable models with boundaries can be
constructed out of a pair of reflection $K$-matrices $K^{\pm }(u)$ in
addition to the solution of the {\small YB} equation. Here $K^{-}(u)$ and $%
K^{+}(u)$ describe the effects of the presence of boundaries at the left and
the right ends, respectively.

Integrability of open chains in the framework of the quantum inverse
scattering method was pioneered by Sklyanin relying on previous results of
Cherednik \cite{Che}. In reference \cite{Skl}, Sklyanin has used his
formalism to solve, via algebraic {\small BA}, the open spin-$1/2$ chain
with diagonal boundary terms. This \ model had already been solved via
coordinate {\small BA }by Alcaraz {\it et al} \cite{Alc}. The Sklyanin
original formalism was extended to more general systems by Mezincescu and
Nepomechie in \cite{MN}.

In this paper we consider the coordinate version of the {\small BA} for the
trigonometric three-state vertex models with a class of boundary terms
derived from diagonal reflection $K$-matrices. These models are well-known
in the literature: the Zamolodchikov-Fateev ({\small ZF}) model or $%
A_{1}^{(1)}$ model \cite{ZF}, the Izergin-Korepin ({\small IK}) model or $%
A_{2}^{(2)}$ model \cite{IK} and two $Z_{2}$-graded models, named the $%
sl(2|1)$\ model and the $osp(1|2)$ model \cite{BS}.

In the context of the coordinate {\small BA}, we propose here a new
parametrization of wavefunctions. This result is important since it allows
us to treat these $19$-vertex models in the same way. Moreover the
coordinate {\small BA} for these three-states models becomes simple as for
two-state models terms \cite{Alc}.

The main goal in this paper is to reveal the common structure of these $19$%
-vertex models with boundary terms which permits us to apply the {\small BA}
method, unifying old and new results.

The paper is organized as follows: We introduce the algebraic tools in
Section $2$. In section $3$, we apply the coordinate {\small BA} method for
a general open chain Hamitonian associated with four $19$-vertex models. In
sections $4$,$5$,$6$ and $7$ the energy eigenspectra and the corresponding
Bethe equations are presented for each model. In section $8$ we discuss
about the graded and non graded solutions for $19$-vertex models. Section $9$
is reserved for the conclusion.

\section{Description of the model}

To determine an integrable vertex model on a lattice it is first necessary
that the bulk vertex weights be specified by an ${\cal R}$-matrix ${\cal R}%
(u)$, where $u$ is the spectral parameter. It acts on the tensor product $%
V^{1}\otimes V^{2}$ for a given vector space $V$ and satisfy a special
system of functional equations, the {\small YB} equation 
\begin{equation}
{\cal R}_{12}(u){\cal R}_{13}(u+v){\cal R}_{23}(v)={\cal R}_{23}(v){\cal R}%
_{13}(u+v){\cal R}_{12}(u),  \label{des.1}
\end{equation}%
in $V^{1}\otimes V^{2}\otimes V^{3}$, where ${\cal R}_{12}={\cal R}\otimes 
{\bf 1}$, ${\cal R}_{23}={\bf 1}\otimes {\cal R}$, etc.

An ${\cal R}$ matrix is said to be regular if it satisfies the property $%
{\cal R}(0)=P$, where $P$ is the permutation matrix in $V^{1}\otimes V^{2}$: 
$P(\left| \alpha \right\rangle \otimes \left| \beta \right\rangle )=\left|
\beta \right\rangle \otimes \left| \alpha \right\rangle $ for $\left| \alpha
\right\rangle ,\left| \beta \right\rangle \in V$. \ In addition, we will
require \cite{MN} that ${\cal R}(u)$ satisfies the following properties%
\begin{eqnarray}
{\rm regularity} &:&{\cal R}_{12}(0)=f(0)^{1/2}P_{12},  \nonumber \\
{\rm unitarity} &:&{\cal R}_{12}(u){\cal R}_{12}^{t_{1}t_{2}}(-u)=f(u), 
\nonumber \\
{\rm PT-symmetry} &:&P_{12}{\cal R}_{12}(u)P_{12}={\cal R}%
_{12}^{t_{1}t_{2}}(u),  \nonumber \\
{\rm cros}\text{sin}{\rm g-symmetry} &:&{\cal R}_{12}(u)=U_{1}{\cal R}%
_{12}^{t_{2}}(-u-\rho )U_{1}^{-1},  \label{des.2}
\end{eqnarray}%
where $f(u)=x_{1}(u)x_{1}(-u)$, $t_{i}$ denotes transposition in the space $%
i $ , $\rho $ is the crossing parameter and $U$ determines the crossing
matrix%
\begin{equation}
M=U^{t}U=M^{t}.  \label{des.3}
\end{equation}%
Note that unitarity and crossing-symmetry together imply the useful relation%
\begin{equation}
M_{1}{\cal R}_{12}^{t_{2}}(-u-\rho )M_{1}^{-1}{\cal R}_{12}^{t_{1}}(u-\rho
)=f(u).  \label{des.4}
\end{equation}

The boundary weights then follow from $K$-matrices which satisfy boundary
versions of the {\small YB} equation \cite{Skl, MN}: the reflection equation%
\begin{equation}
{\cal R}_{12}(u-v)K_{1}^{-}(u){\cal R}%
_{12}^{t_{1}t_{2}}(u+v)K_{2}^{-}(v)=K_{2}^{-}(v){\cal R}%
_{12}(u+v)K_{1}^{-}(u){\cal R}_{12}^{t_{1}t_{2}}(u-v),  \label{des.5}
\end{equation}%
and the dual reflection equation%
\[
{\cal R}_{12}(-u+v)(K_{1}^{+})^{t_{1}}(u)M_{1}^{-1}{\cal R}%
_{12}^{t_{1}t_{2}}(-u-v-2\rho )M_{1}(K_{2}^{+})^{t_{2}}(v) 
\]%
\begin{equation}
=(K_{2}^{+})^{t_{2}}(v)M_{1}{\cal R}_{12}(-u-v-2\rho
)M_{1}^{-1}(K_{1}^{+})^{t_{1}}(u){\cal R}_{12}^{t_{1}t_{2}}(-u+v).
\label{des.6}
\end{equation}%
In this case there is an isomorphism \ between $K^{-}$ and $K^{+}$ :%
\begin{equation}
K^{-}(u):\rightarrow K^{+}(u)=K^{-}(-u-\rho )^{t}M.  \label{des.7}
\end{equation}%
Therefore, given a solution to the reflection equation (\ref{des.5}) we can
also find a solution to the dual reflection equation (\ref{des.6}).

In the framework of the quantum inverse scattering method, we define the Lax
operator from the ${\cal R}$-matrix \ as $L_{aq}(u)={\cal R}_{aq}(u)$, where
the subscript $a$ represents auxiliary space, and $q$ represents quantum
space. The row-to-row monodromy matrix $T(u)$ is defined as a matrix product
over the $N$ \ operators on all sites of the lattice,%
\begin{equation}
T(u)=L_{aN}(u)L_{aN-1}(u)\cdots L_{a1}(u).  \label{des.8}
\end{equation}

The main result is the following: if the boundary equations are satisfied,
then the Sklyanin's transfer matrix%
\begin{equation}
t(u)={\rm Tr}_{a}\left( K^{+}(u)T(u)K^{-}(u)T^{-1}(-u)\right) ,
\label{des.9}
\end{equation}%
forms a commuting family%
\begin{equation}
\left[ t(u),t(v)\right] =0,\qquad \forall u,v  \label{des.10}
\end{equation}

The commutativity of $t(u)$ can be proved by using the unitarity and
crossing-unitarity relations, the reflection equation and the dual
reflection equation. It implies integrability of an open quantum spin chain
whose Hamiltonian (with $K^{-}(0)=1$), can be obtained as%
\begin{equation}
H=\sum_{k=1}^{N-1}H_{k,k+1}+\frac{1}{2}\left. \frac{dK_{1}^{-}(u)}{du}%
\right| _{u=0}+\frac{{\rm tr}_{0}K_{0}^{+}(0)H_{N,0}}{{\rm tr}K^{+}(0)},
\label{des.11}
\end{equation}%
and whose two-site terms are given by%
\begin{equation}
H_{k,k+1}=\left. \frac{d}{du}P_{k,k+1}{\cal R}_{k,k+1}(u)\right| _{u=0},
\label{des.12}
\end{equation}%
in the standard fashion.

Here we will extend our discussions to include the $Z_{2}$-graded vertex
models. Therefore, let us describe some useful informations about the graded
formulation.

Let $V=V_{0}\oplus V_{1}$ be a $Z_{2}$-graded vector space where $0$ and $1$
denote the even and odd parts respectively. Multiplication rules in the
graded tensor product space $V\overset{s}{\otimes }V$ differ from the
ordinary ones by the appearance of additional signs. The components of a
linear operator $A\overset{s}{\otimes }B\in V\overset{s}{\otimes }V$ result
in matrix elements of the form 
\begin{equation}
(A\overset{s}{\otimes }B)_{\alpha \beta }^{\gamma \delta }=(-)^{p(\beta
)(p(\alpha )+p(\gamma ))}\ A_{\alpha \gamma }B_{\beta \delta }.
\label{des.13}
\end{equation}%
The action of the graded permutation operator ${\cal P}$ on the vector $%
\left| \alpha \right\rangle \overset{s}{\otimes }\left| \beta \right\rangle
\in V\overset{s}{\otimes }V$ is defined by 
\begin{equation}
{\cal P}\ \left| \alpha \right\rangle \overset{s}{\otimes }\left| \beta
\right\rangle =(-)^{p(\alpha )p(\beta )}\left| \beta \right\rangle \overset{s%
}{\otimes }\left| \alpha \right\rangle \Longrightarrow ({\cal P})_{\alpha
\beta }^{\gamma \delta }=(-)^{p(\alpha )p(\beta )}\delta _{\alpha \delta }\
\delta _{\beta \gamma }.  \label{des.14}
\end{equation}%
The graded transposition ${\rm st}$ and the graded trace {\rm str} are
defined by 
\begin{equation}
\left( A^{{\rm st}}\right) _{\alpha \beta }=(-)^{(p(\alpha )+1)p(\beta
)}A_{\beta \alpha },\quad {\rm str}A=\sum_{\alpha }(-)^{p(\alpha )}A_{\alpha
\alpha }.  \label{des.15}
\end{equation}%
where $p(\alpha )=1\ (0)$ if $\left| \alpha \right\rangle $ is an odd (even)
element.

For the graded case the {\small YB} equation and the reflection equation
remain the same as above. We only need to change the usual tensor product to
the graded tensor product.

In general, the dual reflection equation which depends on the unitarity and
cross-unitarity relations of the ${\cal R}$-matrix takes different forms for
different models. For the models \ considered in this paper, we write the
graded dual reflection equation in the following form \cite{FW}:%
\[
{\cal R}_{21}^{st_{1}st_{2}}(-u+v)(K_{1}^{+})^{st_{1}}(u)M_{1}^{-1}{\cal R}%
_{12}^{st_{1}st_{2}}(-u-v-2\rho )M_{1}(K_{2}^{+})^{st_{2}}(v) 
\]%
\begin{equation}
=(K_{2}^{+})^{st_{2}}(v)M_{1}{\cal R}_{12}^{st_{1}st_{2}}(-u-v-2\rho
)M_{1}^{-1}(K_{1}^{+})^{st_{1}}(u){\cal R}_{21}^{st_{1}st_{2}}(-u+v),
\label{des.16}
\end{equation}%
and we will choose a common parity assignment: $p(1)=p(3)=0$ \ and $p(2)=1$
, the {\small BFB} grading.

Now, using the relations%
\begin{equation}
{\cal R}_{12}^{st_{1}st_{2}}(u)=I_{1}R_{21}(u)I_{1},\quad {\cal R}%
_{21}^{st_{1}st_{2}}(u)=I_{1}R_{12}(u)I_{1}\quad {\rm and}\quad
IK^{+}(u)I=K^{+}(u)  \label{des.17}
\end{equation}%
with $I={\rm diag}(1,-1,1)$ and the property $\left[ M_{1}M_{2},{\cal R}(u)%
\right] =0$ we can see that the isomorphism (\ref{des.7}) holds with the 
{\small BFB} grading.

The three-state vertex models that we will consider are the
Zamolodchikov-Fateev ({\small ZF}) model, the Izergin-Korepin ({\small IK})
model, the $sl(2|1)$ model and the $osp(1|2)$ model. Their ${\cal R}$%
-matrices have a common form%
\begin{equation}
{\cal R}(u)=\left( 
\begin{array}{ccc|ccc|ccc}
x_{1} &  &  &  &  &  &  &  &  \\ 
& x_{2} &  & x_{5} &  &  &  &  &  \\ 
&  & x_{3} &  & x_{6} &  & x_{7} &  &  \\ \hline
& y_{5} &  & x_{2} &  &  &  &  &  \\ 
&  & y_{6} &  & x_{4} &  & x_{6} &  &  \\ 
&  &  &  &  & x_{2} &  & x_{5} &  \\ \hline
&  &  &  &  &  &  &  &  \\[-10pt] 
&  & y_{7} &  & y_{6} &  & x_{3} &  &  \\ 
&  &  &  &  & y_{5} &  & x_{2} &  \\ 
&  &  &  &  &  &  &  & x_{1}%
\end{array}%
\right) ,  \label{des.18}
\end{equation}%
satisfying the properties (\ref{des.1}--\ref{des.4}) together with their
graded version.

In order to derive the bulk Hamiltonian, it is convenient to expand the
normalized $R$-matrix ($R=P{\cal R}$) around the regular point $u=0$ 
\begin{equation}
R(u,\eta )=1+u(\alpha ^{-1}H+\beta I)+{\rm o}(u^{2}),  \label{des.19}
\end{equation}%
with $\alpha $ and $\beta $ being scalar functions. \ Therefore $H_{k,k+1}$
in (\ref{des.11}) is the $H$ in (\ref{des.19}) acting on the quantum spaces
at sites $k$ and $k+1$.

Using a spin language, this is a {\em spin} $1$ Hamiltonian. In the basis
where $S_{k}^{z}$ is diagonal with eigenvectors $\left| +,k\right\rangle
,\left| 0,k\right\rangle ,\left| -,k\right\rangle $ and eigenvalues $1,0,-1$%
, respectively, the bulk Hamiltonian density acting on two neighboring sites
is given by 
\begin{equation}
H_{k,k+1}=\left( 
\begin{array}{ccc|ccc|ccc}
z_{1} &  &  &  &  &  &  &  &  \\ 
& \overset{\_}{z}_{5} &  & 1 &  &  &  &  &  \\ 
&  & \overset{\_}{z}_{7} &  & \overset{\_}{z}_{6} &  & z_{3} &  &  \\ \hline
& 1 &  & z_{5} &  &  &  &  &  \\ 
&  & \epsilon \overset{\_}{z}_{6} &  & \epsilon z_{4} &  & \epsilon z_{6} & 
&  \\ 
&  &  &  &  & \overset{\_}{z}_{5} &  & 1 &  \\ \hline
&  &  &  &  &  &  &  &  \\[-10pt] 
&  & z_{3} &  & z_{6} &  & z_{7} &  &  \\ 
&  &  &  &  & 1 &  & z_{5} &  \\ 
&  &  &  &  &  &  &  & z_{1}%
\end{array}%
\right) ,  \label{des.20}
\end{equation}%
which can be easily written in terms of the usual spin-$1$ operators: 
\begin{eqnarray}
H_{k,k+1}\! &=&\epsilon z_{4}+\frac{1}{2}(\overset{\_}{z}%
_{5}-z_{5})[S_{k}^{z}-S_{k+1}^{z}]+\frac{1}{2}(z_{5}+\overset{\_}{z}%
_{5}-2\epsilon z_{4})[(S_{k}^{z})^{2}+(S_{k+1}^{z})^{2}]  \nonumber \\
&&+\frac{1}{4}(2z_{1}-z_{7}-\overset{\_}{z}_{7})S_{k}^{z}S_{k+1}^{z}+\frac{1%
}{4}(2z_{1}+z_{7}+\overset{\_}{z}_{7}+4\epsilon z_{4}-4z_{5}-4\overset{\_}{z}%
_{5}))(S_{k}^{z}S_{k+1}^{z})^{2}  \nonumber \\
&&+\frac{1}{4}(z_{7}-\overset{\_}{z}_{7}-2z_{5}+2\overset{\_}{z}%
_{5})[(S_{k}^{z})^{2}S_{k+1}^{z}-S_{k}^{z}(S_{k+1}^{z})^{2}]  \nonumber \\
&&+\frac{1}{4}\epsilon
z_{3}[(S_{k}^{+}S_{k+1}^{-})^{2}+(S_{k}^{-}S_{k+1}^{+})^{2}]  \nonumber \\
&&-\frac{1}{2}[(z_{6}S_{k}^{+}S_{k+1}^{-}+\overset{\_}{z}%
_{6}S_{k}^{-}S_{k+1}^{+})S_{k}^{z}S_{k+1}^{z}+S_{k}^{z}S_{k+1}^{z}(\overset{%
\_}{z}_{6}S_{k}^{+}S_{k+1}^{-}+z_{6}S_{k}^{-}S_{k+1}^{+})]  \nonumber \\
&&+\frac{1}{2}[S_{k}^{+}S_{k}^{z}S_{k+1}^{z}S_{k+1}^{-}\!+%
\!S_{k}^{-}S_{k}^{z}S_{k+1}^{z}S_{k+1}^{+}\!+%
\!S_{k}^{z}S_{k}^{+}S_{k+1}^{-}S_{k+1}^{z}\!+%
\!S_{k}^{z}S_{k}^{-}S_{k+1}^{+}S_{k+1}^{z}].  \label{des.21}
\end{eqnarray}%
Here we have used the sign $\epsilon =\pm 1$ \ to explicitly include the
graded ($\epsilon =-1$) models for which the tensor products in (\ref{des.21}%
) are {\small BFB} graded and 
\begin{equation}
S^{z}=\left( 
\begin{array}{ccc}
1 & 0 & 0 \\ 
0 & 0 & 0 \\ 
0 & 0 & -1%
\end{array}%
\right) ,\quad S^{+}=\sqrt{2}\left( 
\begin{array}{ccc}
0 & 1 & 0 \\ 
0 & 0 & 1 \\ 
0 & 0 & 0%
\end{array}%
\right) ,\quad S^{-}=\sqrt{2}\left( 
\begin{array}{ccc}
0 & 0 & 0 \\ 
1 & 0 & 0 \\ 
0 & 1 & 0%
\end{array}%
\right) .  \label{des.22}
\end{equation}%
Finally, we can consider the boundary terms of $H$ which are derived from
the diagonal solutions $K^{\pm }={\rm diag}(k_{11}^{\pm },k_{22}^{\pm
},k_{33}^{\pm })$ of the reflection equations. In general, at the first site
(left) it has the form%
\begin{equation}
\frac{1}{2}\alpha \frac{dK^{-}(u)}{du}=\left( 
\begin{array}{ccc}
l_{11} &  &  \\ 
& l_{22} &  \\ 
&  & l_{33}%
\end{array}%
\right) ,\quad l_{ii}=\frac{1}{2}\alpha \left. \frac{dk_{ii}^{-}(u)}{du}%
\right| _{u=0},\quad i=1,2,3.  \label{des.23}
\end{equation}%
and for the last site (right) it has the form%
\begin{equation}
\frac{{\rm tr}_{0}K_{0}^{+}(0)H_{N,0}}{{\rm tr}K^{+}(0)}=\left( 
\begin{array}{ccc}
r_{11} &  &  \\ 
& r_{22} &  \\ 
&  & r_{33}%
\end{array}%
\right) .  \label{des.24}
\end{equation}%
To compute the term ${\rm tr}_{0}K_{0}^{+}(0)H_{N,0}$ we have to use (\ref%
{des.21}) in order to get $H_{N,0}$. In practice, it is equivalent to take
the trace of \ $K_{+}(0)H_{21}$ where 
\begin{equation}
H_{21}=P_{12}H_{12}P_{12}=\left( 
\begin{array}{ccc}
h_{11} & h_{12} & h_{13} \\ 
h_{21} & h_{22} & h_{23} \\ 
h_{31} & h_{32} & h_{33}%
\end{array}%
\right) ,  \label{des.25}
\end{equation}%
with $h_{ij}$ being $3$ by $3$ matrices and $H_{12}$ is given by (\ref%
{des.20}). The result for the right boundary term is%
\begin{eqnarray}
r_{11} &=&\frac{z_{1}k_{11}^{+}(0)+\epsilon \overset{\_}{z}_{5}k_{22}^{+}(0)+%
\overset{\_}{z}_{7}k_{33}^{+}(0)}{k_{11}^{+}(0)+\epsilon
k_{22}^{+}(0)+k_{33}^{+}(0)},\quad r_{22}=\frac{%
z_{5}k_{11}^{+}(0)+z_{4}k_{22}^{+}(0)+\overset{\_}{z}_{5}k_{33}^{+}(0)}{%
k_{11}^{+}(0)+\epsilon k_{22}^{+}(0)+k_{33}^{+}(0)},  \nonumber \\
r_{33} &=&\frac{z_{7}k_{11}^{+}(0)+\epsilon
z_{5}k_{22}^{+}(0)+z_{1}k_{33}^{+}(0)}{k_{11}^{+}(0)+\epsilon
k_{22}^{+}(0)+k_{33}^{+}(0)}  \label{des.25a}
\end{eqnarray}

The factor $\alpha $ in (\ref{des.23}) is due to the normalization of $H$ in
(\ref{des.19}) and $\epsilon $'s appear in (\ref{des.25a}) to take into
account the graded traces. Therefore the most general diagonal boundary
terms can be written as%
\begin{eqnarray}
{\rm b.t.} &=&\frac{1}{2}(l_{11}^{^{\prime }}-l_{33}^{^{\prime }})S_{1}^{z}+%
\frac{1}{2}(l_{11}^{^{\prime }}+l_{33}^{^{\prime
}})(S_{1}^{z})^{2}+l_{22}1_{1}  \nonumber \\
&&+\frac{1}{2}(r_{11}^{^{\prime }}-r_{33}^{^{\prime }})S_{N}^{z}+\frac{1}{2}%
(r_{11}^{^{\prime }}+r_{33}^{^{\prime }})(S_{N}^{z})^{2}+r_{22}1_{N},
\label{des.26}
\end{eqnarray}%
where $l_{ii}^{^{\prime }}=l_{ii}-l_{22}$ and \ $r_{ii}^{^{\prime
}}=r_{ii}-r_{22}$ for $i=1,2,3$.

\section{The coordinate Bethe Ansatz}

In this section results are presented for a open quantum spin chain of $N$
atoms each with spin $1$ described by the Hamiltonian (\ref{des.21}) with
the boundaries term (\ref{des.26}): 
\begin{equation}
H=\sum_{k=1}^{N-1}H_{k,k+1}+{\rm b.t.}  \label{cba.1a}
\end{equation}%
At each site, the {\em spin variable} may be $+1,0,-1$, so that the Hilbert
space of the spin chain is $H^{(N)}=\otimes ^{N}V$ where $V=C^{3}$ with
basis $\left\{ \left| +\right\rangle ,\left| 0\right\rangle ,\left|
-\right\rangle \right\} $. The dimension of the Hilbert space is {\rm dim}$%
H^{(N)}=3^{N}$. We can see that $H$ commutes with the third component of the 
{\em spin} 
\begin{equation}
\left[ H,S_{T}^{z}\right] =0,\qquad S_{T}^{z}=\sum_{k=1}^{N}S_{k}^{z},
\label{cba.1}
\end{equation}%
This allows us to divide the Hilbert space of states into different sectors,
each labelled by the eigenvalue of the number operator $r=N-S_{T}^{z}$ . We
shall denote by $H_{n}^{(N)}$ the subspace of $H^{(N)}$ with $r=n$. We can
see that ${\rm dim}H^{(N)}=\sum_{r=0}^{N}{\rm dim}H_{r}^{(N)}$ with 
\begin{equation}
{\rm dim}H_{r}^{(N)}=\sum_{j=0}^{[\frac{r}{2}]}\left( 
\begin{array}{c}
N \\ 
r-2j%
\end{array}%
\right) \left( 
\begin{array}{c}
N-r+2j \\ 
j%
\end{array}%
\right) ,  \label{cba.2}
\end{equation}%
where $[\frac{r}{2}]$ means the integer part of $\frac{r}{2}$ and $\left( 
\begin{array}{c}
a \\ 
b%
\end{array}%
\right) $ denotes the binomial number.

The action of $H_{k,k+1}$ on two neighboring sites is read directly of (\ref%
{des.20})%
\begin{eqnarray}
H_{k,k+1}\left| ++\right\rangle &=&z_{1}\left| ++\right\rangle ,\quad
H_{k,k+1}\left| --\right\rangle =z_{1}\left| --\right\rangle ,  \nonumber \\
\quad H_{k,k+1}\left| +0\right\rangle &=&\overset{\_}{z}_{5}\left|
+0\right\rangle +\left| 0+\right\rangle ,\quad H_{k,k+1}\left|
0+\right\rangle =z_{5}\left| 0+\right\rangle +\left| +0\right\rangle , 
\nonumber \\
H_{k,k+1}\left| 0-\right\rangle &=&\overset{\_}{z}_{5}\left| 0-\right\rangle
+\left| -\ 0\right\rangle ,\quad H_{k,k+1}\left| -\ 0\right\rangle
=z_{5}\left| -\ 0\right\rangle +\left| 0-\right\rangle ,  \nonumber \\
H_{k,k+1}\left| +-\right\rangle &=&\overset{\_}{z}_{7}\left| +-\right\rangle
+\overset{\_}{z}_{6}\left| 0\ 0\right\rangle +z_{3}\left| -+\right\rangle , 
\nonumber \\
H_{k,k+1}\left| 0\ 0\right\rangle &=&\epsilon z_{4}\left| 0\ 0\right\rangle
+\epsilon \overset{\_}{z}_{6}\left| +-\right\rangle +\epsilon z_{6}\left|
-+\right\rangle ,  \nonumber \\
H_{k,k+1}\left| -+\right\rangle &=&z_{7}\left| -+\right\rangle +\
z_{6}\left| 0\ 0\right\rangle +z_{3}\left| +-\right\rangle ,  \label{cba.3}
\end{eqnarray}%
and the boundary terms (\ref{des.26}) only see the sites $1$ and $N$ .
Therefore, we can write 
\begin{equation}
{\rm b.t.}\left| i,...,j\right\rangle ={\cal E}_{ij}\left|
i,...,j\right\rangle ,\quad  \label{cba.4}
\end{equation}%
where $\ {\cal E}_{ij}=l_{ii}+r_{jj},\quad i,j=1,2,3$ , with the notation $%
(+,0,-)=(1,2,3)$. $l_{ii}$ and $r_{jj}$ are given by (\ref{des.23}) and (\ref%
{des.25}), respectively.

\subsection{Sector r=0}

The sector $H_{0}^{(N)}$ contains only one state, the {\em reference state},
with all spin value equal to $+1$, $\Psi _{0}=\prod_{k}\left|
+,k\right\rangle $, satisfying $H\Psi _{0}=E_{0}\Psi _{0}$, with $%
E_{0}=(N-1)z_{1}+{\cal E}_{11}$. All other energies will be measured
relative to this state. It means that we will seek eigenstates of $H$
satisfying $(H-E_{0})\Psi _{r}={\cal E}_{r}\Psi _{r}$ , in every sector $r$.

\subsection{Sector r=1}

In $H_{1}^{(N)}${\it ,} the subspace of states with all spin value equal to $%
+1$ except one with value $0$. There are $N$ states $\left|
k[0]\right\rangle =\left| +++\,{%
\raisebox{-0.65em}{$\stackrel{\textstyle
0}{\scriptstyle k}$}}\,++\cdots +\right\rangle $ which span a basis of $%
H_{1}^{(N)}$. The Ansatz for the eigenstate is thus of the form 
\begin{equation}
\Psi _{1}=\sum_{k=1}^{N}a(k)\left| k[0]\right\rangle .  \label{cba.5}
\end{equation}%
The unknown wavefunction $a(k)$ determines the probability that the {\em %
spin variable} has the value $0$ at the site $k$.

When $H$ acts on $\left| k[0]\right\rangle $ , it sees the reference
configuration, except in the vicinity of $k$, and using (\ref{cba.3}) we
obtain the eigenvalue equations 
\begin{equation}
({\cal E}_{1}+2z_{1}-z_{5}-\overset{\_}{z}_{5})a(k)=a(k-1)+a(k+1),\quad
(1<k<N)  \label{cba.6}
\end{equation}%
At the boundaries, we get slightly different equations%
\begin{eqnarray}
({\cal E}_{1}+{\cal E}_{11}-{\cal E}_{21}+z_{1}-z_{5})a(1) &=&a(2), 
\nonumber \\
({\cal E}_{1}+{\cal E}_{11}-{\cal E}_{12}+z_{1}-\overset{\_}{z}_{5})a(N)
&=&a(N-1).  \label{cba.7}
\end{eqnarray}%
We now try as a solution 
\begin{equation}
a(k)=a(\theta )\xi ^{k}-a(-\theta )\xi ^{-k},  \label{cba.9}
\end{equation}%
where $\xi ={\rm e}^{i\theta }$ , $\theta $ being some particular momentum
fixed by the boundary conditions. Substituting this in equation (\ref{cba.6}%
) we obtain the eigenvalue 
\begin{equation}
{\cal E}_{1}=-2z_{1}+z_{5}+\overset{\_}{z}_{5}+\xi +\xi ^{-1}.
\label{cba.10}
\end{equation}

We want equations (\ref{cba.6}) to be valid for $k=1$ and $k=N$ also, where $%
a(0)$ and $a(N+1)$ are defined by (\ref{cba.9}). Combining (\ref{cba.7})
with (\ref{cba.6}) we get the end conditions%
\begin{equation}
\begin{array}{c}
a(0)=\Delta _{1}a(1),\quad \Delta _{1}={\cal E}_{21}-{\cal E}_{11}+z1-%
\overset{\_}{z}_{5}=z_{1}-\overset{\_}{z}_{5}-l_{11}^{^{\prime }}, \\ 
\\ 
a(N+1)=\Delta _{2}a(N),\quad \Delta _{2}={\cal E}_{12}-{\cal E}%
_{11}+z1-z_{5}=z_{1}-z_{5}-r_{11}^{^{\prime }}.%
\end{array}
\label{cba.11}
\end{equation}%
Compatibility between the end conditions (\ref{cba.11}) yields 
\begin{equation}
\frac{a(\theta )}{a(-\theta )}=\xi ^{-2}\frac{\Delta _{1}-\xi }{\Delta
_{1}-\xi ^{-1}}=\xi ^{-2N}\frac{\Delta _{2}-\xi ^{-1}}{\Delta _{2}-\xi },
\label{cba.14}
\end{equation}%
or%
\begin{equation}
\xi ^{2N}=\left( \frac{\Delta _{1}\xi -1}{\Delta _{1}-\xi }\right) \left( 
\frac{\Delta _{2}\xi -1}{\Delta _{2}-\xi }\right) .  \label{cba.15}
\end{equation}%
Therefore, the energy eigenvalue of $H$ in the sector $\ r=1$ is given by%
\begin{equation}
E_{1}=(N-3)z_{1}+l_{11}+r_{11}+z_{5}+\overset{\_}{z}_{5}+\xi +\xi ^{-1},
\label{cba.16}
\end{equation}%
with $\xi $ being solution of (\ref{cba.15}).

\subsection{Sector r=2}

In the Hilbert space $H_{2}^{(N)}$ we have $N$ states of the type $\left|
k[-]\right\rangle =\left| ++\,{%
\raisebox{-0.65em}{$\stackrel{\textstyle
-}{\scriptstyle k}$}}\,++\cdots +\right\rangle $ and $N(N-1)/2$ states of
the type $\left| k_{1}[0],k_{2}[0]\right\rangle =\left| ++\,{%
\raisebox{-0.65em}{$\stackrel{\textstyle 0}{\scriptstyle k_1}$}}\,++\,{%
\raisebox{-0.65em}{$\stackrel{\textstyle 0}{\scriptstyle k_2}$}}\,++\cdots
+\right\rangle $. We seek these eigenstates in the form 
\begin{equation}
\Psi _{2}=\sum_{k_{1}<k_{2}}a(k_{1},k_{2})\left|
k_{1}[0],k_{2}[0]\right\rangle +\sum_{k=1}^{N}b(k)\left| k[-]\right\rangle .
\label{cba.17}
\end{equation}

Following Bethe \cite{Bethe}, the wavefunction $a(k_{1},k_{2})$ can be
parametrized using the superposition of plane waves (\ref{cba.9}) including
the scattering of two {\em pseudoparticles} with {\em momenta} $\theta _{1}$
and $\theta _{2}$, ($\xi _{j}=e^{i\theta _{j}},j=1,2$): 
\begin{equation}
a(k_{1},k_{2})=\sum_{P}\varepsilon _{P}\left\{ a(\theta _{1},\theta _{2})\xi
_{1}^{k_{1}}\xi _{2}^{k_{2}}-a(\theta _{2},\theta _{1})\xi _{2}^{k_{1}}\xi
_{1}^{k_{2}}\right\} \ ,  \label{cba.18}
\end{equation}%
where the sum extends over the negations of $\theta _{1}$ and $\theta _{2}$,
and $\varepsilon _{P}$ is a sign factor ($\pm 1$) that changes sign on
negation. The parametrization of $b(k)$ is still undetermined at this stage.

Before we try to parametrize $b(k)$ let us consider the Schr\"{o}dinger
equation $H\Psi _{2}=E_{2}\Psi _{2}$ . From the explicit form of $H$ acting
on two sites (\ref{cba.3}) we derive the following set of eigenvalue
equations:

\begin{itemize}
\item Equations for $\left| k_{1}[0]\right\rangle $ and $\left|
k_{2}[0]\right\rangle $ far in the bulk ($1\!<k_{1}\!<k_{2}\!+\!1\!<\!N$) 
\begin{equation}
({\cal E}_{2}\!\!+\!\!4z_{1}\!\!-\!\!2z_{5}\!\!-\!\!2\overset{\_}{z}%
_{5})a(k_{1},k_{2})=a(k_{1}\!\!-1,k_{2})\!+\!a(k_{1}\!\!+1,k_{2})%
\!+a(k_{1},k_{2}\!\!-\!\!1)\!+\!a(k_{1},k_{2}\!+\!1).  \label{cba.19}
\end{equation}

\item Equations for $\left| k[-]\right\rangle $ in the bulk ($1<k<N$) 
\begin{equation}
({\cal E}_{2}+2z_{1}-z_{7}-\overset{\_}{z}_{7})b(k)=z_{3}b(k-1)+z_{3}b(k+1)+%
\overset{\_}{z}_{6}a(k-1,k)+z_{6}a(k,k+1).  \label{cba.20}
\end{equation}

\item Equations for two $\left| k[0]\right\rangle $ neighbors in the bulk ($%
1<k<N-1$) 
\begin{equation}
({\cal E}_{2}+3z_{1}-z_{5}-\overset{\_}{z}_{5}-\epsilon
z_{4})a(k,k+1)=a(k-1,k+1)+a(k,k+2)+\epsilon z_{6}b(k)\!+\epsilon \overset{\_}%
{z}_{6}b(k+1).  \label{cba.21}
\end{equation}
\end{itemize}

In addition we have seven conditions to be satisfied at the free ends of the
chain:

\begin{itemize}
\item Five equations involving at least one state $\left| k[0]\right\rangle $
at one of the ends 
\begin{equation}
({\cal E}_{2}+{\cal E}_{11}\!-\!{\cal E}_{21}+3z_{1}\!-\!2z_{5}\!-\overset{\_%
}{z}_{5})a(1,k_{2})=a(2,k_{2})+a(1,k_{2}\!-1)+a(1,k_{2}\!+1),  \label{cba.22}
\end{equation}%
\begin{equation}
({\cal E}_{2}\!+\!{\cal E}_{11}\!\!-\!\!{\cal E}_{12}\!+3z_{1}\!\!-z_{5}\!%
\!-\!\!2\overset{\_}{z}_{5})a(k_{1},N)=a(k_{1}\!\!-\!\!1,N)+a(k_{1}\!%
\!+1,N)+a(k_{1},N\!\!-1),  \label{cba.23}
\end{equation}%
\begin{equation}
({\cal E}_{2}+{\cal E}_{11}-{\cal E}_{22}+2z_{1}-z_{5}-\overset{\_}{z}%
_{5})a(1,N)=a(2,N)+a(1,N-1),  \label{cba.24}
\end{equation}%
\begin{equation}
({\cal E}_{2}+{\cal E}_{11}-{\cal E}_{21}+2z_{1}-z_{5}-\epsilon
z_{4})a(1,2)=a(1,3)+\epsilon z_{6}b(1)+\epsilon \overset{\_}{z}_{6}b(2),
\label{cba.25}
\end{equation}%
\begin{equation}
({\cal E}_{2}\!+{\cal E}_{11}\!\!-{\cal E}_{12}+2z_{1}\!-\overset{\_}{z}%
_{5}\!-\!\epsilon z_{4})a(N\!-1,N)=a(N\!\!-2,N)+\epsilon
z_{6}b(N\!\!-1)+\epsilon \overset{\_}{z}_{6}b(N).  \label{cba.26}
\end{equation}

\item Two equations with the state $\left| k[-]\right\rangle $ at one of the
ends%
\begin{equation}
({\cal E}_{2}+{\cal E}_{11}-{\cal E}%
_{31}+z_{1}-z_{7})b(1)=z_{3}b(2)+z_{6}a(1,2),  \label{cba.27}
\end{equation}%
\begin{equation}
({\cal E}_{2}+{\cal E}_{11}-{\cal E}_{13}+z_{1}-\overset{\_}{z}%
_{7})b(N)=z_{3}b(N-1)+\overset{\_}{z}_{6}a(N-1,N).  \label{cba.28}
\end{equation}
\end{itemize}

By simples substitution the Ansatz (\ref{cba.18}) solves the equations (\ref%
{cba.19}) provided 
\begin{equation}
{\cal E}_{2}=-4z_{1}+2z_{5}+2\overset{\_}{z}_{5}+\xi _{1}+\xi _{1}^{-1}+\xi
_{2}+\xi _{2}^{-1}.  \label{cba.29}
\end{equation}%
It immediately follows that the eigenvalues of $H$ are a sum of single
pseudoparticle energies.

The parametrization of $b(k)$ can now be determined in the following way:
subtracting Eq.(\ref{cba.21}) from Eq.(\ref{cba.19}) for $k_{1}=k,k_{2}=k+1$%
, we get 
\begin{equation}
\varepsilon \overset{\_}{z}_{6}b(k+1)+\varepsilon
z_{6}b(k)=a(k,k)+a(k+1,k+1)-(z_{1}+\varepsilon z_{4}-z_{5}-\overset{\_}{z}%
_{5})a(k,k+1).  \label{cba.30}
\end{equation}%
for which we can find $b(k)$ in terms of $a(k_{1},k_{2})$.

Using (\ref{cba.30}) together with (\ref{cba.19}) we can see that (\ref%
{cba.25}) and (\ref{cba.26}) are readily satisfied. Now we extend the Ansatz
(\ref{cba.18}) to $k_{1}=k_{2}=k$ in order to get a parametrization for the
wavefunction $b(k)$ :%
\begin{eqnarray}
b(k) &=&\sum_{P}\varepsilon _{P}\!b(\theta _{1},\theta _{2})\xi _{1}^{k}\xi
_{2}^{k}  \nonumber \\
&=&b(\theta _{1},\theta _{2})\xi _{1}^{k}\xi _{2}^{k}\!\!-b(-\theta
_{1},\theta _{2})\xi _{1}^{-k}\xi _{2}^{k}-b(\theta _{1},\!-\theta _{2})\xi
_{1}^{k}\xi _{2}^{-k}+b(-\theta _{1},\!-\theta _{2})\xi _{1}^{-k}\xi
_{2}^{-k},  \label{cba.31}
\end{eqnarray}%
which solves the meeting condition (\ref{cba.30}) provided 
\begin{eqnarray}
\epsilon b(\theta _{1},\theta _{2}) &=&\left( \frac{1+\xi _{1}\xi
_{2}+\Delta \xi _{2}}{z_{6}+\overset{\_}{z}_{6}\xi _{1}\xi _{2}}\right)
a(\theta _{1},\theta _{2})-\left( \frac{1+\xi _{1}\xi _{2}+\Delta \xi _{1}}{%
z_{6}+\overset{\_}{z}_{6}\xi _{1}\xi _{2}}\right) a(\theta _{2},\theta _{1}),
\nonumber \\
\Delta &=&z_{5}+\overset{\_}{z}_{5}-z_{1}-\varepsilon z_{4},  \label{cba.32}
\end{eqnarray}%
together with $b(-\theta _{1},\theta _{2}),b(\theta _{1},-\theta _{2})$ and $%
b(-\theta _{1},-\theta _{2})$ that can be obtained from (\ref{cba.32})
changing the signs of $\theta _{1}$ and $\theta _{2}$ .

These relations tell us that the pseudoparticle of the type $\left|
k[-]\right\rangle $ behaves under the action of $H$ as the two
pseudoparticles $\left| k_{1}[0]\right\rangle $ and $\left|
k_{2}[0]\right\rangle $ at the same site $k$ and its parametrization follows
as the plane waves of pseudoparticles $\left| k_{i}[0]\right\rangle $
multiplied by the weight functions $b(\pm \theta _{1},\pm \theta _{2})$.

As a consequence of this identification we can see that the equation
involving $b(k)$ (\ref{cba.20}) becomes a {\em meeting condition} for two
states $\left| k[0]\right\rangle $. Using the $S$-matrix language, from (\ref%
{cba.20}) we get the {\em two-pseudopartcle phase shifts}:%
\begin{eqnarray}
a(\theta _{2},\theta _{1}) &=&\left( \frac{s(\theta _{2},\theta _{1})}{%
s(\theta _{1},\theta _{2})}\right) a(\theta _{1},\theta _{2}),\quad a(\theta
_{2},-\theta _{1})=\left( \frac{s(\theta _{2},-\theta _{1})}{s(-\theta
_{1},\theta _{2})}\right) a(-\theta _{1},\theta _{2}),  \nonumber \\
a(-\theta _{2},\theta _{1}) &=&\left( \frac{s(-\theta _{2},\theta _{1})}{%
s(\theta _{1},-\theta _{2})}\right) a(\theta _{1},-\theta _{2}),\quad
a(-\theta _{2},-\theta _{1})=\left( \frac{s(-\theta _{2},-\theta _{1})}{%
s(-\theta _{1},-\theta _{2})}\right) a(-\theta _{1},-\theta _{2}),
\label{cba.33}
\end{eqnarray}%
where 
\begin{eqnarray}
s(\theta _{2},\theta _{1}) &=&\left( 1+\xi _{1}\xi _{2}+\Delta \xi
_{2}\right) \left[ z_{3}(1+\xi _{1}^{2}\xi _{2}^{2})-(1+\xi _{1}\xi
_{2})(\xi _{1}+\xi _{2})+\Lambda \xi _{1}\xi _{2}\right]  \nonumber \\
&&+\epsilon \xi _{2}\left( z_{6}+\overset{\_}{z}_{6}\xi _{1}\xi _{2}\right)
\left( \overset{\_}{z}_{6}+z_{6}\xi _{1}\xi _{2}\right) ,  \nonumber \\
\Lambda &=&2(z_{1}-z_{5}-\overset{\_}{z}_{5})+z_{7}+\overset{\_}{z}_{7}.
\label{cba.34}
\end{eqnarray}%
The seven remained $s$-functions for the phase shift equations (\ref{cba.33}%
) follow from (\ref{cba.34}) changing the signs of $\theta _{1}$ and $\theta
_{2}$.

At this point we still have to consider\ the equation (\ref{cba.24}) and
four end conditions. We want equation (\ref{cba.19}) to be valid for $%
k_{1}=1 $ and $k_{2}=N$ also, where $a(0,k_{2})$ and $a(k_{1},N+1)$ are
defined by (\ref{cba.18}). Combining (\ref{cba.22}) and (\ref{cba.23}) with (%
\ref{cba.19}) we get two end conditions%
\begin{equation}
\begin{array}{c}
\Delta _{1}a(1,k)=a(0,k),\quad \Delta _{1}=z_{1}-\overset{\_}{z}_{5}-{\cal E}%
_{11}+{\cal E}_{21,} \\ 
\\ 
\Delta _{2}a(k,N)=a(k,N+1),\quad \Delta _{2}=z_{1}-z_{5}-{\cal E}_{11}+{\cal %
E}_{12}%
\end{array}
\label{cba.35}
\end{equation}%
Substituting (\ref{cba.18}) in (\ref{cba.35}), we obtain the following
relations 
\begin{eqnarray}
a(-\theta _{1},\theta _{2}) &=&\left( \frac{1-\Delta _{1}\xi _{1}}{1-\Delta
_{1}\xi _{1}^{-1}}\right) a(\theta _{1},\theta _{2}),\quad a(\theta
_{1},-\theta _{2})=\xi _{2}^{2N}\left( \frac{\Delta _{2}-\xi _{2}}{\Delta
_{2}-\xi _{2}^{-1}}\right) a(\theta _{1},\theta _{2}),  \nonumber \\
a(-\theta _{1},-\theta _{2}) &=&\xi _{2}^{2N}\left( \frac{1-\Delta _{1}\xi
_{1}}{1-\Delta _{1}\xi _{1}^{-1}}\right) \left( \frac{\Delta _{2}-\xi _{2}}{%
\Delta _{2}-\xi _{2}^{-1}}\right) a(\theta _{1},\theta _{2}).  \label{cba.39}
\end{eqnarray}%
which describe the change of signs of $\theta _{1}$ and $\theta _{2}$ in $%
a(\theta _{1},\theta _{2})$ and the corresponding pair interchange relations%
\begin{eqnarray}
a(\theta _{2},-\theta _{1}) &=&\xi _{1}^{2N}\left( \frac{\Delta _{2}-\xi _{1}%
}{\Delta _{2}-\xi _{1}^{-1}}\right) a(\theta _{2},\theta _{1}),\quad
a(-\theta _{2},\theta _{1})=\left( \frac{1-\Delta _{1}\xi _{2}}{1-\Delta
_{1}\xi _{2}^{-1}}\right) a(\theta _{2},\theta _{1}),  \nonumber \\
a(-\theta _{2},-\theta _{1}) &=&\xi _{1}^{2N}\left( \frac{1-\Delta _{1}\xi
_{2}}{1-\Delta _{1}\xi _{2}^{-1}}\right) \left( \frac{\Delta _{2}-\xi _{1}}{%
\Delta _{2}-\xi _{1}^{-1}}\right) a(\theta _{2},\theta _{1}).  \label{cba.40}
\end{eqnarray}%
Combining these relations with the phase shift relations (\ref{cba.33}) we
get the Bethe equations%
\begin{eqnarray}
\xi _{1}^{2N} &=&\left( \frac{1-\Delta _{1}\xi _{1}}{\Delta _{1}-\xi _{1}}%
\right) \left( \frac{1-\Delta _{2}\xi _{1}}{\Delta _{2}-\xi _{1}}\right)
\left( \frac{s(\theta _{1},\theta _{2})}{s(\theta _{2},\theta _{1})}\right)
\left( \frac{s(\theta _{2},-\theta _{1})}{s(-\theta _{1},\theta _{2})}%
\right) ,  \label{cba.41} \\
&&  \nonumber \\
\xi _{2}^{2N} &=&\left( \frac{1-\Delta _{1}\xi _{2}}{\Delta _{1}-\xi _{2}}%
\right) \left( \frac{1-\Delta _{2}\xi _{2}}{\Delta _{2}-\xi _{2}}\right)
\left( \frac{s(\theta _{2},\theta _{1})}{s(\theta _{1},\theta _{2})}\right)
\left( \frac{s(\theta _{1},-\theta _{2})}{s(-\theta _{2},\theta _{1})}%
\right) .  \label{cba.42}
\end{eqnarray}%
With these relations we have defined the behavior of the state $\left|
k[0]\right\rangle $ at the boundaries. Consequently the equation (\ref%
{cba.24}), where we have one state $\left| k[0]\right\rangle $ at each end,
is also readily satisfied.

Now we recall step by step our procedure described above to see that there
is no more function or parameter to be determined, but we still have to
solve the equations with the pseudoparticle $\left| -\right\rangle $ at the
boundaries.

Similarly, we want equation (\ref{cba.20}) to be valid for $k=1$ and $k=N$
also, where $b(0)$ and $b(N+1)$ are defined by (\ref{cba.31}). Combining (%
\ref{cba.27}) and (\ref{cba.28}) with (\ref{cba.20}) we obtain two further
end conditions%
\begin{equation}
\begin{array}{c}
\Delta _{3}b(1)=z_{3}b(0)+\overset{\_}{z}_{6}a(0,1),\quad \Delta _{3}=z_{1}-%
\overset{\_}{z}_{7}-{\cal E}_{11}+{\cal E}_{31}, \\ 
\\ 
\Delta _{4}b(N)=z_{3}b(N+1)+z_{6}a(N,N+1),\quad \Delta _{4}=z_{1}-z_{7}-%
{\cal E}_{11}+{\cal E}_{13}.%
\end{array}
\label{cba.43}
\end{equation}%
Substituting the already fixed relations for $a(k_{1},k_{2})$ and $b(k)$ and
using the Bethe equations we can see that these end conditions are also
satisfied. It means that there is no additional end condition due to the
presence of the state $\left| k[-]\right\rangle $\ at the boundaries. \
Instead of surprising, this result is in agreement with our parametrization
of the wavefunction $b(k)$, where the dynamics of the pseudoparticle $\left|
k[-]\right\rangle $ is understood as the dynamics of two pseudoparticles $%
\left| k[0]\right\rangle $. Therefore, the energy eigenvalue for the sector $%
r=2$ is%
\begin{equation}
E_{2}=(N-1)z_{1}+l_{11}+r_{11}+\sum_{j=1}^{2}\left( -2z_{1}+z_{5}+\overset{\_%
}{z}_{5}+\xi _{j}+\xi _{j}^{-1}\right) ,  \label{cba.45}
\end{equation}%
with%
\begin{eqnarray}
\xi _{j}^{2N} &=&\left( \frac{1-\Delta _{1}\xi _{j}}{\Delta _{1}-\xi _{j}}%
\right) \left( \frac{1-\Delta _{2}\xi _{j}}{\Delta _{2}-\xi _{j}}\right)
\prod_{k=1,\ k\neq j}^{2}\left( \frac{s(\theta _{j},\theta _{k})}{s(\theta
_{k},\theta _{j})}\right) \left( \frac{s(\theta _{k},-\theta _{j})}{%
s(-\theta _{j},\theta _{k})}\right) ,\   \nonumber \\
j &=&1,2.
\end{eqnarray}%
where the functions $s(\theta _{j},\theta _{k})$ are given by (\ref{cba.34}).

\subsection{General sector}

The above results can now be generalized to arbitrary values of $r$. \ In a
generic sector $r$ we build eigenstates of $H$ as direct products of $N_{0}$
states $\left| k[0]\right\rangle $ and $N_{-}$ states $\left|
k[-]\right\rangle $, such that $r=N_{0}+2N_{-}$ . These eigenstates are
obtained by superposition of terms of the form 
\begin{equation}
\left| \phi _{r}\right\rangle =\left| 0\right\rangle \times \left| \phi
_{r-1}\right\rangle +\left| -\right\rangle \times \left| \phi
_{r-2}\right\rangle ,  \label{cba.47}
\end{equation}%
with $\left| \phi _{0}\right\rangle =1$, $\left| \phi _{1}\right\rangle
=\left| 0\right\rangle $. For instance, in the sector $r=3$ the eigenstate
of $H$ \ has the form%
\begin{eqnarray}
\Psi _{3} &=&\sum_{k_{1}<k_{2}<k_{3}}a(k_{1},k_{2},k_{3})\left|
k_{1}[0],k_{2}[0],k_{3}[0]\right\rangle  \nonumber \\
&&+\sum_{k_{1}<k_{2}}\left\{ b_{1}(k_{1},k_{2})\left|
k_{1}[-],k_{2}[0]\right\rangle +b_{2}(k_{1},k_{2})\left|
k_{1}[0],k_{2}[-]\right\rangle \right\} .  \label{cba.48}
\end{eqnarray}

The Ansatz for the wavefunction of the term with $N_{0}$ states $\left|
k[0]\right\rangle $ becomes%
\begin{equation}
a(k_{1},k_{2},\ldots ,k_{r})=\sum_{P}\varepsilon _{p}a(\theta _{1},\theta
_{2},\ldots ,\theta _{r})\xi _{1}^{k_{1}}\xi _{2}^{k_{2}}\cdots \xi
_{r}^{k_{r}},  \label{cba.49}
\end{equation}%
where the sum extends over all permutations and negations of $\theta
_{1},\theta _{2},\ldots ,\theta _{r}$ and $\varepsilon _{p}$ changes sign at
each such mutation.

The Ansatz for the wavefunction of terms with $N_{-}$ \ states $\left|
k[-]\right\rangle $ follows from (\ref{cba.49}) as sum over negations of the
terms with $2N_{-}$ \ states $\left| k[0]\right\rangle $ at the same site.
For instance, in the sector $r=3$%
\begin{eqnarray}
b_{1}(k_{1},k_{2}) &=&\left( \sum_{P}\varepsilon _{P}b_{11}(\theta
_{1},\theta _{2})\xi _{1}^{k_{1}}\xi _{2}^{k_{1}}\right) \xi
_{3}^{k_{2}}+\left( \sum_{P}\varepsilon _{P}b_{12}(\theta _{1},\theta
_{3})\xi _{1}^{k_{1}}\xi _{3}^{k_{1}}\right) \xi _{2}^{k_{2}}  \nonumber \\
&&+\left( \sum_{P}\varepsilon _{P}b_{13}(\theta _{2},\theta _{3})\xi
_{1}^{k_{1}}\xi _{2}^{k_{1}}\right) \xi _{1}^{k_{2}},  \label{cba.50}
\end{eqnarray}%
with similar equation for $b_{2}(k_{1},k_{2})$. In that way we always have a
\ {\em far} pseudoparticle $\left| k[0]\right\rangle $ as a {\em viewer}. We
also have verified\ in this sector that the meeting of $\left|
k[0]\right\rangle $ with $\left| k[-]\right\rangle $ can be versed as a
meeting of three $\left| k[0]\right\rangle $ whose phase shift factorizes in
a product of \ two-pseudoparticle phase shifts.

The corresponding energy eigenvalue is a sum of single one-particle energies%
\begin{equation}
E_{r}=(N-1)z_{1}+l_{11}+r_{11}+\sum_{j=1}^{r}\left( -2z_{1}+z_{5}+\overset{\_%
}{z}_{5}+\xi _{j}+\xi _{j}^{-1}\right) ,  \label{cba.51}
\end{equation}%
where $\xi _{j}$ are solutions of the Bethe equations 
\begin{eqnarray}
\xi _{j}^{2N} &=&\left( \frac{1-\Delta _{1}\xi _{j}}{\Delta _{1}-\xi _{j}}%
\right) \left( \frac{1-\Delta _{2}\xi _{j}}{\Delta _{2}-\xi _{j}}\right)
\prod_{k=1,\ k\neq j}^{r}\left( \frac{s(\theta _{j},\theta _{k})}{s(\theta
_{k},\theta _{j})}\right) \left( \frac{s(\theta _{k},-\theta _{j})}{%
s(-\theta _{j},\theta _{k})}\right) ,  \nonumber \\
j &=&1,...,r  \label{cba.52}
\end{eqnarray}%
with%
\begin{eqnarray}
s(\theta _{j},\theta _{k}) &=&\left( 1+\xi _{j}\xi _{k}+\Delta \xi
_{j}\right) \left[ z_{3}(1+\xi _{j}^{2}\xi _{k}^{2})-(1+\xi _{j}\xi
_{k})(\xi _{j}+\xi _{k})+\Lambda \xi _{j}\xi _{k}\right]  \nonumber \\
&&+\epsilon \xi _{j}\left( z_{6}+\overset{\_}{z}_{6}\xi _{j}\xi _{k}\right)
\left( \overset{\_}{z}_{6}+z_{6}\xi _{j}\xi _{k}\right) ,  \label{cba.53}
\end{eqnarray}%
and%
\begin{eqnarray}
\Lambda &=&2(z_{1}-z_{5}-\overset{\_}{z}_{5})+z_{7}+\overset{\_}{z}_{7}, 
\nonumber \\
\Delta _{1} &=&z_{1}-\overset{\_}{z}_{5}-l_{11}+l_{22},\quad \Delta
_{2}=z_{1}-z_{5}-r_{11}+r_{22}.  \label{cba.54}
\end{eqnarray}

Having now built a common ground for all open spin-$1$ Hamiltonians
associated with the 19-vertex models, we may proceed to find explicitly
their spectra. We will do that in the next sections.

\section{The Zamolodchikov-Fateev model}

The simplest three-states vertex model is the {\small ZF} $19$-vertex \cite%
{ZF} or the $A_{1}^{(1)}$ model the spin-$1$ representation \cite{KRS} and
can be constructed from the six-vertex model using the fusion procedure. The 
${\cal R}$-matrix \ which satisfies the {\small YB} equation (\ref{des.1})
has the form (\ref{des.18}) with%
\begin{eqnarray}
x_{1}(u) &=&\sinh (u+\eta )\sinh (u+2\eta ),\quad x_{2}(u)=\sinh u\sinh
(u+\eta ),  \nonumber \\
x_{3}(u) &=&\sinh u\sinh (u-\eta ),\quad x_{4}(u)=\sinh u\sinh (u+\eta
)+\sinh \eta \sinh 2\eta ,  \nonumber \\
y_{5}(u) &=&x_{5}(u)=\sinh (u+\eta )\sinh 2\eta ,\quad
y_{6}(u)=x_{6}(u)=\sinh u\sinh 2\eta ,  \nonumber \\
y_{7}(u) &=&x_{7}(u)=\sinh \eta \sinh 2\eta .  \label{zf.1}
\end{eqnarray}%
This ${\cal R}$-matrix is regular and unitary, with $f(u)=x_{1}(u)x_{1}(-u)$%
, $P$- and $T$-symmetric and crossing-symmetric with $M=1$ and $\rho =\eta $%
. The most general diagonal solution for $K^{-}(u)$ has been obtained in
Ref. \cite{Mez} and is given by%
\begin{equation}
K^{-}(u,\beta _{11})=\left( 
\begin{array}{ccc}
k_{11}^{-}(u) &  &  \\ 
& 1 &  \\ 
&  & k_{33}^{-}(u)%
\end{array}%
\right) ,  \label{zf.2}
\end{equation}%
with%
\begin{equation}
k_{11}^{-}(u)=-\frac{\beta _{11}\sinh u+2\cosh u}{\beta _{11}\sinh u-2\cosh u%
},\quad k_{33}^{-}(u)=-\frac{\beta _{11}\sinh (u+\eta )-2\cosh (u+\eta )}{%
\beta _{11}\sinh (u-\eta )+2\cosh (u-\eta )},  \label{zf.3}
\end{equation}%
where $\beta _{11}$ \ is the free parameter. By the automorphism (\ref{des.7}%
) the solution for $K^{+}(u)$ follows 
\begin{equation}
K^{+}(u,\alpha _{11})=K^{-}(-u-\rho ,\alpha _{11})=\left( 
\begin{array}{ccc}
k_{11}^{+}(u) &  &  \\ 
& 1 &  \\ 
&  & k_{33}^{+}(u)%
\end{array}%
\right) ,  \label{zf.4}
\end{equation}%
with%
\begin{equation}
k_{11}^{+}(u)=-\frac{\alpha _{11}\sinh (u+\eta )-2\cosh (u+\eta )}{\alpha
_{11}\sinh (u+\eta )+2\cosh (u+\eta )},\quad k_{33}^{+}(u)=-\frac{\alpha
_{11}\sinh u+2\cosh u}{\alpha _{11}\sinh (u+2\eta )-2\cosh (u+2\eta )},
\label{zf.5}
\end{equation}%
where $\alpha _{11}$ is another free parameter.

We recall section $2$ to derive the corresponding quantum open spin chain
Hamiltonian. It is the quantum open spin chain for the spin-1 {\small XXZ}
model. 
\begin{equation}
H=\sum_{k=1}^{N-1}H_{k,k+1}+{\rm b.t.}  \label{zf.6}
\end{equation}%
where the bulk Hamiltonian is given by (\ref{des.21}) with the weights%
\begin{eqnarray}
\epsilon &=&1,\quad \alpha =\sinh 2\eta ,\quad z_{1}=0,\quad z_{3}=-1,\quad
z_{4}=-2\cosh 2\eta ,  \nonumber \\
\overset{\_}{z}_{5} &=&z_{5}=-\cosh 2\eta ,\quad \overset{\_}{z}%
_{6}=z_{6}=2\cosh \eta ,\quad \overset{\_}{z}_{7}=z_{7}=-1-2\cosh 2\eta .
\label{zf.7}
\end{eqnarray}%
\ and the boundary terms given by (\ref{des.26}). For the left boundary (\ref%
{des.23}) we get 
\begin{equation}
l_{11}=\frac{1}{2}\beta _{11}\sinh 2\eta ,\quad l_{22}=0,\quad l_{33}=\frac{%
\beta _{11}\cosh \eta -2\sinh \eta }{\beta _{11}\sinh \eta -2\cosh \eta }%
\sinh 2\eta ,  \label{zf.8}
\end{equation}%
and%
\begin{eqnarray}
r_{11}-r_{22} &=&\frac{\alpha _{11}\cosh \eta -2\sinh \eta }{\alpha
_{11}\sinh \eta -2\cosh \eta }\sinh 2\eta ,\quad r_{33}-r_{22}=\frac{1}{2}%
\alpha _{11}\sinh 2\eta ,  \nonumber \\
r_{22} &=&-\frac{1}{4}\frac{\sinh 4\eta }{\sinh 3\eta }\left( \frac{[\alpha
_{11}\sinh \eta +2\cosh \eta ]^{2}-4[1+2\cosh 2\eta ]}{\alpha _{11}\sinh
\eta -2\cosh \eta }\right) ,  \label{zf.9}
\end{eqnarray}%
for the right boundary (\ref{des.25a}).

Next, we can use the coordinate {\small BA} method as described in section $%
3 $ to find the energy eigenvalue (\ref{cba.51}) and the {\small BA}
equations (\ref{cba.52}). Here we just list the results. The energy spectrum
of the Hamiltonian (\ref{zf.6}) for a generic sector $r$ is given by%
\begin{equation}
E_{r}=\sinh 2\eta \left( \frac{1}{2}\beta _{11}+\frac{\alpha _{11}\cosh \eta
-2\sinh \eta }{\alpha _{11}\sinh \eta -2\cosh \eta }\right)
+r_{22}+\sum_{j=1}^{r}(-2\cosh 2\eta +\xi _{j}+\xi _{j}^{-1}),  \label{zf.10}
\end{equation}%
with $\xi _{j}={\rm e}^{i\theta _{j}}$\ satisfying the Bethe equations%
\begin{eqnarray}
\xi _{j}^{2N} &=&\left( \frac{\Delta _{1}\xi _{j}-1}{\Delta _{1}-\xi _{j}}%
\right) \left( \frac{\Delta _{2}\xi _{j}-1}{\Delta _{2}-\xi _{j}}\right)
\prod_{k\neq j}^{n}\left( \frac{s(\theta _{j},\theta _{k})}{s(\theta
_{k},\theta _{j})}\right) \left( \frac{s(\theta _{k},-\theta _{j})}{%
s(-\theta _{j},\theta _{k})}\right) ,  \nonumber \\
j &=&1,2,...,n  \label{zf.11}
\end{eqnarray}%
where%
\begin{equation}
\Delta _{1}=\sinh 2\eta \left( \coth 2\eta -\frac{1}{2}\beta _{11}\right)
,\quad \Delta _{2}=\sinh 2\eta \left( \coth 2\eta -\frac{\alpha _{11}\cosh
\eta -2\sinh \eta }{\alpha _{11}\sinh \eta -2\cosh \eta }\right) .
\label{zf.12}
\end{equation}%
For the {\small ZF} model the two-particle phase shift is given by%
\begin{equation}
\frac{s(\theta _{j},\theta _{k})}{s(\theta _{k},\theta _{j})}=\frac{1+\xi
_{j}+\xi _{k}+\xi _{j}\xi _{k}-(\Delta +2)\xi _{j}}{1+\xi _{j}+\xi _{k}+\xi
_{j}\xi _{k}-(\Delta +2)\xi _{k}},  \label{zf.13}
\end{equation}%
with the $s$-functions given by (\ref{cba.53}) and $\Delta =2\cosh 2\eta $.

This energy spectrum was already obtained\ by Mezincescu {\it at al.} \cite%
{Mez} through a generalization of the quantum inverse scattering method
developed by Sklyanin \cite{Skl}, the so-called fusion procedure \cite{KRS}.
This fusion procedure was also used byYung and Batchelor \cite{YB} to solve
the {\small ZF} vertex-model with inhomogeneities.

For a particular choice of boundary terms, the {\small ZF} spin chain has
the quantum group symmetry {\it i.e.}, if we choose $\xi _{\mp }\rightarrow
\infty $ \ ($\beta _{11}=2\coth \xi _{-}$ \ and $\ \alpha _{11}=2\coth \xi
_{+}$), then the spin chain Hamiltonian (\ref{zf.1}) has $U_{q}(su(2))$%
-invariance \cite{Mez}.

\section{The Izergin-Korepin model}

The solution of the {\small YB} equation corresponding to $A_{2}^{(2)}$ in
the fundamental representation was found by Izergin and Korepin \cite{IK}.
The ${\cal R}$-matrix has the form (\ref{des.18}) with non-zero entries%
\begin{eqnarray}
x_{1}(u) &=&\sinh (u-5\eta )+\sinh \eta ,\quad x_{2}(u)=\sinh (u-3\eta
)+\sinh 3\eta ,  \nonumber \\
x_{3}(u) &=&\sinh (u-\eta )+\sinh \eta ,\quad x_{4}(u)=\sinh (u-3\eta
)-\sinh 5\eta +\sinh 3\eta +\sinh \eta  \nonumber \\
x_{5}(u) &=&-2{\rm e}^{-u/2}\sinh 2\eta \cosh (\frac{u}{2}-3\eta ),\quad
y_{5}(u)=-2{\rm e}^{u/2}\sinh 2\eta \cosh (\frac{u}{2}-3\eta )  \nonumber \\
x_{6}(u) &=&2{\rm e}^{-u/2+2\eta }\sinh 2\eta \sinh (\frac{u}{2}),\quad
y_{6}(u)=-2{\rm e}^{u/2-2\eta }\sinh 2\eta \sinh (\frac{u}{2})  \nonumber \\
x_{7}(u) &=&-2{\rm e}^{-u+2\eta }\sinh \eta \sinh 2\eta -{\rm e}^{-\eta
}\sinh 4\eta ,\quad  \nonumber \\
y_{7}(u) &=&2{\rm e}^{u-2\eta }\sinh \eta \sinh 2\eta -{\rm e}^{\eta }\sinh
4\eta .
\end{eqnarray}%
This ${\cal R}$-matrix is regular and unitary, with $f(u)=x_{1}(u)x_{1}(-u)$%
. It is PT-symmetric and crossing-symmetric, with $\rho =-6\eta -i\pi $ and%
\begin{equation}
M=\left( 
\begin{array}{ccc}
{\rm e}^{2\eta } &  &  \\ 
& 1 &  \\ 
&  & {\rm e}^{-2\eta }%
\end{array}%
\right) .  \label{ik.2}
\end{equation}

Diagonal solutions for $K^{-}(u)$ have been obtained in \cite{MN2}. It turns
out that there are three solutions without free parameters, being $K^{-}(u)=1
$, $K^{-}(u)=F^{+}$ and $K^{-}(u)=F^{-}$, with%
\begin{equation}
F^{\pm }=\left( 
\begin{array}{ccc}
{\rm e}^{-u}f^{(\pm )}(u) &  &  \\ 
& 1 &  \\ 
&  & {\rm e}^{u}f^{(\pm )}(u)%
\end{array}%
\right) ,  \label{ik.3}
\end{equation}%
where we have defined%
\begin{equation}
f^{(\pm )}(u)=\frac{\cosh (u/2-3\eta )\pm i\sinh (u/2)}{\cosh (u/2+3\eta
)\mp i\sinh (u/2)}.  \label{ik.4}
\end{equation}%
By the automorphism (\ref{des.7}), three solutions $K^{+}(u)$ follow as $%
K^{+}(u)=M$, $K^{+}(u)=G^{+}$ and $K^{+}(u)=G^{-}$, with%
\begin{equation}
G^{\pm }=\left( 
\begin{array}{ccc}
{\rm e}^{u-4\eta }g^{(\pm )}(u) &  &  \\ 
& 1 &  \\ 
&  & {\rm e}^{-u+4\eta }g^{(\pm )}(u)%
\end{array}%
\right) ,  \label{ik.5}
\end{equation}%
where we have defined%
\begin{equation}
g^{(\pm )}(u)=\frac{\cosh (u/2-3\eta )\pm i\sinh (u/2)}{\cosh (u/2-3\eta
)\mp i\sinh (u/2-6\eta )}.  \label{ik.6}
\end{equation}

The corresponding quantum open spin chain Hamiltonians is also written as in
(\ref{zf.6}), where the bulk term is given by (\ref{des.21}) with

\begin{eqnarray}
\epsilon &=&1,\quad \alpha =-2\sinh 2\eta ,\quad z_{1}=0,\quad z_{3}=\frac{%
\cosh \eta }{\cosh 3\eta },\quad z_{4}=-2\frac{\sinh \eta \sinh 4\eta }{%
\cosh 3\eta }  \nonumber \\
z_{5} &=&-{\rm e}^{-2\eta },\quad \overset{\_}{z}_{5}=-{\rm e}^{2\eta
},\quad z_{6}={\rm e}^{2\eta }\frac{\sinh 2\eta }{\cosh 3\eta },\quad 
\overset{\_}{z}_{6}=-{\rm e}^{-2\eta }\frac{\sinh 2\eta }{\cosh 3\eta } 
\nonumber \\
z_{7} &=&-\frac{\cosh \eta }{\cosh 3\eta }\left( {\rm e}^{-4\eta }+2\sinh
2\eta \right) ,\quad \overset{\_}{z}_{7}=-\frac{\cosh \eta }{\cosh 3\eta }%
\left( {\rm e}^{4\eta }-2\sinh 2\eta \right) .  \label{ik.8}
\end{eqnarray}%
To derive the boundary term (\ref{des.26}), we will only consider three
types of boundary solutions, one for each pair ($K^{-}(u),K^{+}(u)$) defined
by the automorphism (\ref{des.7}): $(1,M)$, $(F^{+},G^{+})$ and $%
(F^{-},G^{-})$.

{\bf The }$(1,M)${\bf \ case: }For $K^{-}(u)=1$ the left boundary (\ref%
{des.23}) vanishes ( $l_{11}=l_{22}=l_{33}=0$ ), and the right boundary (\ref%
{des.24}) is proportional to the identity. To see this we can substitute (%
\ref{ik.8}) and $K^{+}(0)={\rm diag}({\rm e}^{2\eta },1,{\rm e}^{-2\eta })$
in (\ref{des.25a}) to get%
\begin{equation}
r_{11}=r_{22}=r_{33}=-2\frac{\cosh 4\eta \sinh 2\eta }{\sinh 6\eta }.
\label{ik.9}
\end{equation}%
Therefore, the corresponding open chain Hamiltonian is%
\begin{equation}
H=\sum_{k=1}^{N-1}H_{k,k+1}-2\frac{\cosh 4\eta \sinh 2\eta }{\sinh 6\eta }%
1_{N}  \label{ik.10}
\end{equation}

The coordinate {\small BA} gives us the corresponding energy spectrum (\ref%
{cba.51}). \ For a given sector $r$ it is given by%
\begin{equation}
E_{r}=-2\frac{\cosh 4\eta \sinh 2\eta }{\sinh 6\eta }+\sum_{j=1}^{r}\left(
-2\cosh 2\eta +\xi _{j}+\xi _{j}^{-1}\right) ,  \label{ik.11}
\end{equation}%
where $\xi _{j}$ $={\rm e}^{i\theta _{j}}$ are solutions of the Bethe
equations 
\begin{eqnarray}
\xi _{j}^{2N} &=&\prod_{k=1,\ k\neq j}^{r}\left( \frac{s(\theta _{j},\theta
_{k})}{s(\theta _{k},\theta _{j})}\right) \left( \frac{s(\theta _{k},-\theta
_{j})}{s(-\theta _{j},\theta _{k})}\right) ,  \nonumber \\
j &=&1,...,r  \label{ik.12}
\end{eqnarray}%
The two-particle phase shift for the {\small IK} model is given by%
\begin{equation}
\frac{s(\theta _{j},\theta _{k})}{s(\theta _{k},\theta _{j})}=\left( \frac{%
1+\xi _{j}\xi _{k}-\Delta \xi _{j}}{1+\xi _{j}\xi _{k}-\Delta \xi _{k}}%
\right) \left( \frac{1+\xi _{j}\xi _{k}-\xi _{j}-\xi _{k}-(\Delta -2)\xi _{k}%
}{1+\xi _{j}\xi _{k}-\xi _{j}-\xi _{k}-(\Delta -2)\xi _{j}}\right) ,
\label{ik.13}
\end{equation}%
where the $s$-functions are given by (\ref{cba.53}) and $\Delta =2\cosh
2\eta $.

It was noted by Mezincescu and Nepomechie \cite{MN2}, that this open spin
chain Hamiltonian is quantum group invariant. Moreover, the corresponding
transfer matrix has been diagonalized by the analytic {\small BA} in \cite%
{MN3}, using the $U_{q}(su(2))$ invariance of (\ref{ik.10}). This quantum
group invariance was also used by Yung and Batchelor \cite{YB} to determine
properties of the transfer matrix eigenvalues with inhomogeneities essential
to apply the analytical {\small BA}.

{\bf The }$(F^{+},G^{+})${\bf \ and }$(F^{-},G^{-})${\bf \ cases: }These
cases can be treated simultaneously. The matrix elements of the left
boundary term are given by

\begin{equation}
l_{11}^{(\pm )}=\sinh 2\eta \left( \frac{{\rm e}^{3\eta }\mp i}{\cosh 3\eta }%
\right) ,\quad l_{22}^{(\pm )}=0,\quad l_{33}^{(\pm )}=-\sinh 2\eta \left( 
\frac{{\rm e}^{-3\eta }\pm i}{\cosh 3\eta }\right)  \label{ik.15}
\end{equation}%
and for the right boundary term we have%
\begin{eqnarray}
r_{11}^{(\pm )}-r_{22}^{(\pm )} &=&-\sinh 2\eta \left( \frac{{\rm e}^{-3\eta
}\pm i}{\cosh 3\eta }\right) ,\quad r_{33}^{(\pm )}-r_{22}^{(\pm )}=\sinh
2\eta \left( \frac{{\rm e}^{3\eta }\mp i}{\cosh 3\eta }\right) ,  \nonumber
\\
r_{22}^{(\pm )} &=&-\frac{\sinh 4\eta }{\sinh 6\eta }\left( \frac{\cosh
7\eta \pm 4i\sinh 3\eta \sinh \eta \sinh 2\eta }{\cosh 3\eta \pm i\sinh
2\eta }\right) .  \label{ik.16}
\end{eqnarray}%
In these cases the corresponding open chain Hamiltonians are not $%
U_{q}(su(2))$-invariant \cite{MN3}. Nevertheless, it has recently been
argued by Nepomechie \cite{Nepo} that the transfer matrices corresponding to
these solutions also have the $U_{q}(o(3))$ symmetry, but with a nonstandard
coproduct. They can be written in the following form%
\begin{equation}
H^{(\pm )}=\sum_{k=1}^{N-1}H_{k,k+1}+\sinh 2\eta \left(
S_{1}^{z}-S_{N}^{z}+\left( \frac{\sinh 3\eta \mp i}{\cosh 3\eta }\right) %
\left[ (S_{1}^{z})^{2}+(S_{N}^{z})^{2}\right] \right) +r_{22}^{(\pm )}1_{N}.
\label{ik.17}
\end{equation}%
From the coordinate {\small BA} we have find their energy spectra (\ref%
{cba.51}):%
\begin{equation}
E_{r}^{(\pm )}=2\sinh 2\eta \left( \frac{\sinh 3\eta \mp i}{\cosh 3\eta }%
\right) +r_{22}^{(\pm )}+\sum_{j=1}^{r}\left( -2\cosh 2\eta +\xi _{j}+\xi
_{j}^{-1}\right) ,  \label{ik.18}
\end{equation}%
where $\xi _{j}$ $={\rm e}^{i\theta _{j}}$ are solutions of the Bethe
equations 
\begin{eqnarray}
\left( \xi _{j}^{(\pm )}\right) ^{2N} &=&\left( \frac{1-\Delta _{1}^{(\pm
)}\xi _{j}}{\Delta _{1}^{(\pm )}-\xi _{j}}\right) \left( \frac{1-\Delta
_{2}^{(\pm )}\xi _{j}}{\Delta _{2}^{(\pm )}-\xi _{j}}\right) \prod_{k=1,\
k\neq j}^{r}\left( \frac{s(\theta _{j},\theta _{k})}{s(\theta _{k},\theta
_{j})}\right) \left( \frac{s(\theta _{k},-\theta _{j})}{s(-\theta
_{j},\theta _{k})}\right) ,  \nonumber \\
j &=&1,...,r  \label{ik.19}
\end{eqnarray}%
where%
\begin{equation}
\Delta _{1}^{(\pm )}={\rm e}^{2\eta }-\sinh 2\eta \left( \frac{{\rm e}%
^{3\eta }\mp i}{\cosh 3\eta }\right) ,\quad \Delta _{2}^{(\pm )}={\rm e}%
^{-2\eta }+\sinh 2\eta \left( \frac{{\rm e}^{-3\eta }\pm i}{\cosh 3\eta }%
\right) ,\quad  \label{ik.20}
\end{equation}%
and the two-particle phase shift is still given by (\ref{ik.13}). These
cases were also considered in Ref.\cite{YB} through the analytical {\small BA%
} with inhomogeneities.

Finally, we note that is interesting to reformulate the Bolthzmann weights
of the {\small IK} model by the following transformation%
\begin{equation}
{\cal R}(u,\eta )\rightarrow {\cal R}^{^{\prime }}(u,\eta )=\frac{1}{2i}%
{\cal R}(2u,-\eta -i\frac{\pi }{2}).  \label{ik.21}
\end{equation}%
This ${\cal R}^{^{\prime }}$ matrix differs from the one given in \cite{MAR}
by a gauge transformation. It is regular and unitary, with $f^{^{\prime
}}(u)=x_{1}^{^{\prime }}(u)x_{1}^{^{\prime }}(-u)$, $PT$-symmetric and
crossing-unitarity with $M^{^{\prime }}={\rm diag}(-{\rm e}^{-2\eta },1,-%
{\rm e}^{2\eta })$ and \ $\rho ^{^{\prime }}=3\eta $. \ After the gauge
transformation ${\cal R}_{12}^{^{\prime \prime }}(u)=V_{1}{\cal R}%
_{12}^{^{\prime }}(u)V_{1}^{-1}$ with $V={\rm diag}({\rm e}^{-u},1,{\rm e}%
^{u})$, can see that $M^{^{\prime \prime }}={\rm diag}(-{\rm e}^{4\eta },1,-%
{\rm e}^{-4\eta })$ and \ $\rho ^{^{\prime \prime }}=\rho ^{^{\prime }}$. In
this case the solution $(F^{+},G^{+})$ can be written as 
\begin{equation}
F^{^{\prime \prime }-}={\rm diag}(1,-\frac{\sinh (u-\frac{3}{2}\eta )}{\sinh
(u+\frac{3}{2}\eta )},1),\quad G^{^{\prime \prime }+}=-{\rm diag}({\rm e}%
^{4\eta },\frac{\sinh (u+\frac{9}{2}\eta )}{\sinh (u+\frac{3}{2}\eta )},{\rm %
e}^{-4\eta }).  \label{ik.22}
\end{equation}%
This solution was used by Fan in \cite{FAN} to find the spectrum of the
corresponding transfer matrix using the algebraic {\small BA} for one and
two-particle excited states.

\section{The $sl(2|1)$-model}

The solution of the graded {\small YB} equation \ corresponding to $sl(2|1)$
\ in the fundamental representation has the form (\ref{des.18}) with
non-zero entries \cite{KS2, BS}:%
\begin{eqnarray}
x_{1}(u) &=&\cosh (u+\eta )\sinh (u+2\eta ),\quad x_{2}(u)=\sinh u\cosh
(u+\eta ),  \nonumber \\
x_{3}(u) &=&\sinh u\cosh (u-\eta ),\quad x_{4}(u)=\sinh u\cosh (u+\eta
)-\sinh 2\eta \cosh \eta .  \nonumber \\
y_{5}(u) &=&x_{5}(u)=\sinh 2\eta \cosh (u+\eta ),\quad
y_{6}(u)=x_{6}(u)=\sinh 2\eta \sinh u,  \nonumber \\
y_{7}(u) &=&x_{7}(u)=\sinh 2\eta \cosh \eta .  \label{sl.1}
\end{eqnarray}%
This ${\cal R}$-matrix is regular and unitary, with $f(u)=x_{1}(u)x_{1}(-u)$%
, $P$- and $T$-symmetric and crossing-symmetric with $M=1$ and $\rho =\eta $%
. The graded version of the crossing-unitarity relation (\ref{des.4}) is
satisfied with $f^{^{\prime }}(u)=x_{1}(u+i\frac{\pi }{2})x_{1}(-u-i\frac{%
\pi }{2}).$

The most general diagonal solution for $K^{-}(u)$ has been presented in Ref. 
\cite{MN4} and it is given by

\begin{equation}
K^{-}(u,\beta _{11})=\left( 
\begin{array}{ccc}
k_{11}^{-}(u) &  &  \\ 
& 1 &  \\ 
&  & k_{33}^{-}(u)%
\end{array}%
\right) ,  \label{sl.2}
\end{equation}%
with%
\begin{equation}
k_{11}^{-}(u)=-\frac{\beta _{11}\sinh u+2\cosh u}{\beta _{11}\sinh u-2\cosh u%
},\quad k_{33}^{-}(u)=\frac{\beta _{11}\cosh (u+\eta )-2\sinh (u+\eta )}{%
\beta _{11}\cosh (u-\eta )+2\sinh (u-\eta )},  \label{sl.3}
\end{equation}%
where $\beta _{11}$ \ is the free parameter. Due to the automorphism (\ref%
{des.7}) the solution for $K^{+}(u)$ is given by $K^{-}(-u-\rho ,\frac{1}{4}%
\alpha _{11})$ {\it i.e}.%
\begin{equation}
K^{+}(u,\beta _{11})=\left( 
\begin{array}{ccc}
k_{11}^{+}(u) &  &  \\ 
& 1 &  \\ 
&  & k_{33}^{+}(u)%
\end{array}%
\right) ,  \label{sl.4}
\end{equation}%
where%
\begin{equation}
k_{11}^{+}(u)=\frac{\alpha _{11}\cosh (u+\eta )-2\sinh (u+\eta )}{\alpha
_{11}\cosh (u+\eta )+2\sinh (u+\eta )},\quad k_{33}^{+}(u)=-\frac{\alpha
_{11}\sinh u+2\cosh u}{\alpha _{11}\sinh (u+2\eta )-2\cosh (u+2\eta )},
\label{sl.5}
\end{equation}%
and $\alpha _{11}$ is another free parameter.

The weights for the corresponding bulk Hamiltonian (\ref{des.21}) are given
by 
\begin{eqnarray}
\epsilon &=&-1,\quad \alpha =\sinh 2\eta ,\quad z_{1}=0,\quad z_{3}=1,\quad
z_{4}=2\cosh 2\eta ,  \nonumber \\
\overset{\_}{z}_{5} &=&z_{5}=-\cosh 2\eta ,\quad \overset{\_}{z}%
_{6}=z_{6}=2\sinh \eta ,\quad \overset{\_}{z}_{7}=z_{7}=1-2\cosh 2\eta
\label{sl.6}
\end{eqnarray}%
The left boundary terms of {\rm b.t.} (\ref{des.26}) are given by%
\begin{equation}
l_{11}=\frac{1}{2}\beta _{11}\sinh 2\eta ,\quad l_{22}=0,\quad l_{33}=\frac{%
\beta _{11}\sinh \eta -2\cosh \eta }{\beta _{11}\cosh \eta -2\sinh \eta }%
\sinh 2\eta  \label{sl.7}
\end{equation}%
and for the right boundary we have%
\begin{eqnarray}
r_{11}-r_{22} &=&\frac{\alpha _{11}\sinh \eta -2\cosh \eta }{\alpha
_{11}\cosh \eta -2\sinh \eta }\sinh 2\eta ,\quad r_{33}-r_{22}=\frac{1}{2}%
\alpha _{11}\sinh 2\eta ,  \nonumber \\
r_{22} &=&-\frac{1}{4}\frac{\sinh 4\eta }{\cosh 3\eta }\left( \frac{(\alpha
_{11}\cosh \eta +2\sinh \eta )^{2}+4(1-2\cosh 2\eta )}{\alpha _{11}\cosh
\eta -2\sinh \eta }\right) .  \label{sl.8}
\end{eqnarray}

Now, using the coordinate {\small BA} we find the energy spectrum (\ref%
{cba.51}) for the $sl(2|1)$ open chain Hamiltonian: 
\begin{equation}
E_{r}=\sinh 2\eta \left( \frac{1}{2}\beta _{11}+\frac{\alpha _{11}\sinh \eta
-2\cosh \eta }{\alpha _{11}\cosh \eta -2\sinh \eta }\right)
+r_{22}+\sum_{j=1}^{r}(-2\cosh 2\eta +\xi _{j}+\xi _{j}^{-1})  \label{sl.9}
\end{equation}%
with $\xi _{j}={\rm e}^{i\theta _{j}}$\ satisfying the Bethe equations%
\begin{eqnarray}
\xi _{j}^{2N} &=&\left( \frac{\Delta _{1}\xi _{j}-1}{\Delta _{1}-\xi _{j}}%
\right) \left( \frac{\Delta _{2}\xi _{j}-1}{\Delta _{2}-\xi _{j}}\right)
\prod_{k\neq j}^{n}\left( \frac{s(\theta _{j},\theta _{k})}{s(\theta
_{k},\theta _{j})}\right) \left( \frac{s(\theta _{k},-\theta _{j})}{%
s(-\theta _{j},\theta _{k})}\right)  \nonumber \\
j &=&1,2,...,n  \label{sl.10}
\end{eqnarray}%
where%
\begin{equation}
\Delta _{1}=\sinh 2\eta \left( \coth 2\eta -\frac{1}{2}\beta _{11}\right)
,\quad \Delta _{2}=\sinh 2\eta \left( \coth 2\eta -\frac{\alpha _{11}\sinh
\eta -2\cosh \eta }{\alpha _{11}\cosh \eta -2\sinh \eta }\right)
\label{sl.11}
\end{equation}%
and the two-body phase shift for the $sl(2|1)$ model is given by%
\begin{equation}
\frac{s(\theta _{j},\theta _{k})}{s(\theta _{k},\theta _{j})}=\frac{1-\xi
_{j}-\xi _{k}+\xi _{j}\xi _{k}-(\Delta -2)\xi _{j}}{1-\xi _{j}-\xi _{k}+\xi
_{j}\xi _{k}-(\Delta -2)\xi _{k}}  \label{sl.12}
\end{equation}%
where $\Delta =2\cosh 2\eta $.

\section{The $osp(1|2)$-model}

The trigonometric solution of the graded {\small YB} equation \
corresponding to $osp(1|2)$ \ in the fundamental representation has the form
(\ref{des.18}) with non-zero entries \cite{BS}:%
\begin{eqnarray}
x_{1}(u) &=&\sinh (u+2\eta )\sinh (u+3\eta ),\quad x_{2}(u)=\sinh u\sinh
(u+3\eta )  \nonumber \\
x_{3}(u) &=&\sinh u\sinh (u+\eta ),\quad x_{4}(u)=\sinh u\sinh (u+3\eta
)-\sinh 2\eta \sinh 3\eta   \nonumber \\
x_{5}(u) &=&{\rm e}^{-u}\sinh 2\eta \sinh (u+3\eta ),\quad y_{5}(u)={\rm e}%
^{u}\sinh 2\eta \sinh (u+3\eta )  \nonumber \\
x_{6}(u) &=&-{\rm e}^{-u-2\eta }\sinh 2\eta \sinh u,\quad y_{6}(u)={\rm e}%
^{u+2\eta }\sinh 2\eta \sinh u  \nonumber \\
x_{7}(u) &=&{\rm e}^{-u}\sinh 2\eta \left( \sinh (u+3\eta )+{\rm e}^{-\eta
}\sinh u\right)   \nonumber \\
y_{7}(u) &=&{\rm e}^{u}\sinh 2\eta \left( \sinh (u+3\eta )+{\rm e}^{\eta
}\sinh u\right)   \label{osp.1}
\end{eqnarray}%
This ${\cal R}$-matrix is regular and unitary, with $f(u)=x_{1}(u)x_{1}(-u)$%
. It is $PT$-symmetric and crossing-symmetric, with\ $\rho =3\eta $ and%
\begin{equation}
M=\left( 
\begin{array}{ccc}
{\rm e}^{-2\eta } &  &  \\ 
& 1 &  \\ 
&  & {\rm e}^{2\eta }%
\end{array}%
\right) .  \label{osp.2}
\end{equation}

Diagonal solutions for $K^{-}(u)$ have been obtained in \cite{LS}. It turns
out that there are three solutions without free parameters, being $%
K^{-}(u)=1 $, $K^{-}(u)=F^{+}$ and $K^{-}(u)=F^{-}$, with%
\begin{equation}
F^{\pm }=\left( 
\begin{array}{ccc}
\mp {\rm e}^{-2u}f^{(\pm )}(u) &  &  \\ 
& 1 &  \\ 
&  & \mp {\rm e}^{2u}f^{(\pm )}(u)%
\end{array}%
\right) ,  \label{osp.3}
\end{equation}%
where we have defined%
\begin{equation}
f^{(+)}(u)=\frac{\sinh (u+3\eta /2)}{\sinh (u-3\eta /2)},\quad f^{(-)}(u)=%
\frac{\cosh (u+3\eta /2)}{\cosh (u-3\eta /2)}.  \label{osp.4}
\end{equation}%
By the automorphism (\ref{des.7}), three solutions $K^{+}(u)$ follow as $%
K^{+}(u)=M$, $K^{+}(u)=G^{+}$ and $K^{+}(u)=G^{-}$, with%
\begin{equation}
G^{\pm }=\left( 
\begin{array}{ccc}
\mp {\rm e}^{2u+4\eta }g^{(\pm )}(u) &  &  \\ 
& 1 &  \\ 
&  & \mp {\rm e}^{-2u-4\eta }g^{(\pm )}(u)%
\end{array}%
\right) ,  \label{osp.5}
\end{equation}%
where we have defined%
\begin{equation}
g^{(+)}(u)=\frac{\sinh (u+3\eta /2)}{\sinh (u+9\eta /2)},\quad g^{(-)}(u)=%
\frac{\cosh (u+3\eta /2)}{\cosh (u+9\eta /2)}.  \label{osp.6}
\end{equation}

The corresponding quantum open spin chain Hamiltonians is also written as in
(\ref{zf.6}), where the bulk term is given by (\ref{des.21}) with

\begin{eqnarray}
\epsilon &=&-1,\quad \alpha =\sinh 2\eta ,\quad z_{1}=0,\quad z_{3}=\frac{%
\sinh \eta }{\sinh 3\eta },\quad z_{4}=2\frac{\cosh \eta \sinh 4\eta }{\sinh
3\eta }  \nonumber \\
z_{5} &=&-{\rm e}^{2\eta },\quad \overset{\_}{z}_{5}=-{\rm e}^{-2\eta
},\quad z_{6}=-{\rm e}^{-2\eta }\frac{\sinh 2\eta }{\sinh 3\eta },\quad 
\overset{\_}{z}_{6}={\rm e}^{2\eta }\frac{\sinh 2\eta }{\sinh 3\eta }, 
\nonumber \\
z_{7} &=&-{\rm e}^{2\eta }+{\rm e}^{-\eta }\frac{\sinh 2\eta }{\sinh 3\eta }%
,\quad \overset{\_}{z}_{7}=-{\rm e}^{-2\eta }+{\rm e}^{\eta }\frac{\sinh
2\eta }{\sinh 3\eta }  \label{osp.8}
\end{eqnarray}%
To derive the boundary term (\ref{des.26}), we will only consider three
types of boundary solutions, one for each pair ($K^{-}(u),K^{+}(u)$) defined
by the automorphism (\ref{des.7}): $(1,M)$, $(F^{+},G^{+})$ and $%
(F^{-},G^{-})$.

{\bf The }$(1,M)${\bf \ case: }For $K^{-}(u)=1$ the left boundary (\ref%
{des.23}) vanishes ( $l_{11}=l_{22}=l_{33}=0$ ), and the right boundary (\ref%
{des.24}) is proportional to the identity, for which quantum-algebra
invariance is achieved \cite{MN3}. To see this we can substitute (\ref{osp.8}%
) and $K^{+}(0)={\rm diag}({\rm e}^{-2\eta },1,{\rm e}^{2\eta })$ in (\ref%
{des.25a}) to get%
\begin{equation}
r_{11}=r_{22}=r_{33}=2\frac{\cosh 4\eta \sinh 2\eta }{\sinh 6\eta }.
\label{osp.9}
\end{equation}%
Therefore, the corresponding open chain Hamiltonian is%
\begin{equation}
H=\sum_{k=1}^{N-1}H_{k,k+1}+2\frac{\cosh 4\eta \sinh 2\eta }{\sinh 6\eta }%
1_{N}  \label{osp.10}
\end{equation}

The coordinate {\small BA} gives us the corresponding energy spectrum (\ref%
{cba.51}). \ For a given sector $r$ it is given by%
\begin{equation}
E_{r}=2\frac{\cosh 4\eta \sinh 2\eta }{\sinh 6\eta }+\sum_{j=1}^{r}\left(
-2\cosh 2\eta +\xi _{j}+\xi _{j}^{-1}\right)   \label{osp.11}
\end{equation}%
where $\xi _{j}$ $={\rm e}^{i\theta _{j}}$ are solutions of the Bethe
equations 
\begin{eqnarray}
\xi _{j}^{2N} &=&\prod_{k=1,\ k\neq j}^{r}\left( \frac{s(\theta _{j},\theta
_{k})}{s(\theta _{k},\theta _{j})}\right) \left( \frac{s(\theta _{k},-\theta
_{j})}{s(-\theta _{j},\theta _{k})}\right) ,  \nonumber \\
j &=&1,...,r  \label{osp.12}
\end{eqnarray}%
The two-particle phase shift for the $osp(1|2)$ model is given by%
\begin{equation}
\frac{s(\theta _{j},\theta _{k})}{s(\theta _{k},\theta _{j})}=\left( \frac{%
1+\xi _{j}\xi _{k}-\Delta \xi _{j}}{1+\xi _{j}\xi _{k}-\Delta \xi _{k}}%
\right) \left( \frac{1+\xi _{j}\xi _{k}+\xi _{j}+\xi _{k}-(\Delta +2)\xi _{k}%
}{1+\xi _{j}\xi _{k}+\xi _{j}+\xi _{k}-(\Delta +2)\xi _{j}}\right) ,
\label{osp.13}
\end{equation}%
where the $s$-functions are given by (\ref{cba.53}) and $\Delta =2\cosh
2\eta $.

{\bf The }$(F^{+},G^{+})${\bf \ case: \ }In this case the boundary terms are 
\begin{equation}
l_{11}=\frac{{\rm e}^{-3\eta /2}}{\sinh (3\eta /2)}\sinh 2\eta ,\quad
l_{22}=0,\quad l_{33}=\frac{{\rm e}^{3\eta /2}}{\sinh (3\eta /2)}\sinh 2\eta
\label{osp.14}
\end{equation}%
and

\begin{eqnarray}
r_{11}-r_{22} &=&\frac{{\rm e}^{3\eta /2}}{\sinh (3\eta /2)}\sinh 2\eta
,\quad r_{33}-r_{22}=\frac{{\rm e}^{-3\eta /2}}{\sinh (3\eta /2)}\sinh 2\eta
\nonumber \\
r_{22} &=&-\frac{\sinh 4\eta }{\sinh 6\eta }\left( 4\cosh (\frac{3}{2}\eta
)\cosh (\frac{5}{2}\eta )-1\right)  \label{osp.15}
\end{eqnarray}%
The energy eigenvalues are%
\begin{equation}
E_{r}=2\sin 2\eta \coth (3\eta /2)+r_{22}+\sum_{j=1}^{r}\left( -2\cosh 2\eta
+\xi _{j}+\xi _{j}^{-1}\right)  \label{osp.16}
\end{equation}%
with the Bethe equations%
\begin{eqnarray}
\xi _{j}^{2N} &=&\left( \frac{1-\Delta _{1}\xi _{j}}{\Delta _{1}-\xi _{j}}%
\right) \left( \frac{1-\Delta _{2}\xi _{j}}{\Delta _{2}-\xi _{j}}\right)
\prod_{k=1,\ k\neq j}^{r}\left( \frac{s(\theta _{j},\theta _{k})}{s(\theta
_{k},\theta _{j})}\right) \left( \frac{s(\theta _{k},-\theta _{j})}{%
s(-\theta _{j},\theta _{k})}\right) ,  \nonumber \\
j &=&1,...,r  \label{osp.17}
\end{eqnarray}%
where the phase shift is given \ by (\ref{osp.13}) and$\Delta _{1}={\rm e}%
^{-2\eta }-\frac{{\rm e}^{-3\eta /2}}{\sinh (3\eta /2)}\sinh 2\eta ,\quad
\Delta _{2}={\rm e}^{2\eta }-\frac{{\rm e}^{3\eta /2}}{\sinh (3\eta /2)}%
\sinh 2\eta $

{\bf The }${\bf (F}^{-}{\bf ,G}^{-}{\bf )}${\bf \ case: }In this case the
boundary terms are

\begin{eqnarray}
l_{11} &=&-\frac{{\rm e}^{-3\eta /2}}{\cosh (3\eta /2)}\sinh 2\eta ,\quad
l_{22}=0,\quad l_{33}=\frac{{\rm e}^{3\eta /2}}{\cosh (3\eta /2)}\sinh \eta 
\nonumber \\
r_{11}-r_{22} &=&\frac{{\rm e}^{3\eta /2}}{\cosh (3\eta /2)}\sinh 2\eta
,\quad r_{33}-r_{22}=-\frac{{\rm e}^{-3\eta /2}}{\cosh (3\eta /2)}\sinh 2\eta
\nonumber \\
r_{22} &=&-\frac{\sinh 4\eta }{\sinh 6\eta }\left( 4\sinh (\frac{3}{2}\eta
)\sinh (\frac{5}{2}\eta )-1\right)  \label{osp.18}
\end{eqnarray}%
The corresponding energy eigenvalues are given by%
\begin{equation}
E_{r}=2\sin 2\eta \tanh (3\eta /2)+r_{22}+\sum_{j=1}^{r}\left( -2\cosh 2\eta
+\xi _{j}+\xi _{j}^{-1}\right)  \label{osp.19}
\end{equation}%
The Bethe equations are%
\begin{eqnarray}
\xi _{j}^{2N} &=&\left( \frac{1-\Delta _{1}\xi _{j}}{\Delta _{1}-\xi _{j}}%
\right) \left( \frac{1-\Delta _{2}\xi _{j}}{\Delta _{2}-\xi _{j}}\right)
\prod_{k=1,\ k\neq j}^{r}\left( \frac{s(\theta _{j},\theta _{k})}{s(\theta
_{k},\theta _{j})}\right) \left( \frac{s(\theta _{k},-\theta _{j})}{%
s(-\theta _{j},\theta _{k})}\right) ,  \nonumber \\
j &=&1,...,r  \label{osp.20}
\end{eqnarray}%
with the phase shift (\ref{osp.13}) and%
\begin{equation}
\Delta _{1}={\rm e}^{-2\eta }+\frac{{\rm e}^{-3\eta /2}}{\cosh (3\eta /2)}%
,\quad \Delta _{2}={\rm e}^{2\eta }-\frac{{\rm e}^{3\eta /2}}{\cosh (3\eta
/2)}\sinh \eta  \label{osp.21}
\end{equation}

\section{From non graded to graded solutions}

Beside the ${\cal R}$-matrix we also have considered the $R$-matrix, which
satisfies%
\begin{equation}
R_{12}(u)R_{23}(u+v)R_{12}(v)=R_{23}(v)R_{12}(u+v)R_{23}(u).  \label{gra.1}
\end{equation}%
Because only $R_{12}$ and $R_{23}$ are involved, this equation written in
components looks the same as in the non graded case. Moreover, the matrix $%
{\cal R}=PR$ satisfies the usual {\small YB} equation (\ref{des.1}) where $P$
is the non graded permutation matrix. When the graded permutation matrix $%
{\cal P}$\ is used, then ${\cal R}={\cal P}R$ satisfies the graded version
of the {\small YB} equation.

Multiplying the ${\cal R}$-matrix for $19$-vertex models (\ref{des.18}) by
the diagonal matrix $\ \Pi =P{\cal P}={\cal P}P$ we will get graded ${\cal R}
$-matrices starting from non graded ${\cal R}$-matrices and vice-versa. The
new ${\cal R}$-matrix ${\cal R}^{^{\prime }}=\Pi {\cal R}$,\ still has the
form (\ref{des.18}) but with the change of sign of the fifth row due to the
grading {\small BFB}. The bulk Hamiltonian has the form (\ref{des.20}) but
interchanging the role of the sign $\epsilon $. Now $\epsilon =-1$ for
non-graded models and $\epsilon =1$ for graded models.

Let us use this interchange property with the {\small YB} solution of the 
{\small IK} model. First we recall the transformation (\ref{ik.21}) 
\begin{equation}
{\cal R}^{^{\prime }}(u,\eta )=\frac{1}{2i}{\cal R}(2u,-\eta -i\frac{\pi }{2}%
)\Rightarrow H_{k,k+1}^{^{\prime }}(\eta )=H_{k,k+1}(-\eta -i\frac{\pi }{2}).
\label{gra.2}
\end{equation}%
The matrix ${\cal R}_{{\small IKg}}(u,\eta )=\Pi {\cal R}^{^{\prime }}$ is a
solution of the graded version of the {\small YB} equation (\ref{des.1}) and
the corresponding vertex model can be named as the graded version of the 
{\small IK }model.

Using the symmetries of the {\small YB} solutions for $19$-vertex models: $%
x_{2}(u)\rightarrow \pm x_{2}(u)$ and $x_{6}(u)\rightarrow \pm x_{6}(u)$
with $y_{6}(u)\rightarrow \mp y_{6}(u)$, we can see that this model has the
same Boltzmann weights of the $osp(1|2)$-model, except for the presence of
the factor $\pm i$ \ in $x_{6}(u)$ \ and $\mp i$ \ \ in $y_{6}(u)$. However,
this identification is not so trivial due to the change the signs of the
fifth row of ${\cal R}$ ({\small BFB} grading). Nevertheless, by direct
computation we have verified that both models have the same reflection $K$%
-matrices. It means that ${\cal R}_{{\small IKg}}(u,\eta )$ and the ${\cal R}%
(u,\eta )$ of the $osp(1|2)$ share the same symmetries. Consequently, both
open chain Hamiltonians have the same boundary terms. Moreover, from the
definition (\ref{cba.34}) we can see that phase shift equations (\ref{cba.33}%
) are invariant under the replacement\ $z_{6}\rightarrow \pm iz_{6}$ with $%
\overset{\_}{z}_{6}\rightarrow \mp i\overset{\_}{z}_{6}$. Thus, the
coordinate {\small BA} previously described, yields the same spectrum for
both models. In words, the open spin chain Hamiltonians associated with the
graded {\small IK} model have the $osp(1|2)$ - invariance.

This situation is also present in the graded version of the {\small ZF}
model. In order to see that we have to reformulate conveniently the
Boltzmann weights of the {\small ZF} model by the following transformation%
\begin{equation}
{\cal R}(u,\eta )\rightarrow {\cal R}^{^{\prime }}(u,\eta )=\frac{1}{i}{\cal %
R}(u,\eta -i\frac{\pi }{2}).  \label{gra.3}
\end{equation}%
The graded version of the {\small ZF} model is defined by the following $%
{\cal R}$-matrix%
\begin{equation}
{\cal R}_{ZFg}(u,\eta )=\Pi {\cal R}^{^{\prime }}(u,\eta )  \label{gra.4}
\end{equation}%
Using again the symmetries of the $19$-vertex model we can see, up to a
possible canonical transformation: $x_{6}\rightarrow x_{6}^{^{\prime }}(u)$ =%
$\pm ix_{6}(u)$, the non-zero entries of ${\cal R}_{ZFg}(u,\eta )$ are
identified with the Boltzmann weights of the $sl(2|1)$ model (\ref{sl.1}). \
We also find that both models have the same $K$-matrices and their
coordinate Bethe ans\"{a}tze yield a common spectrum.

We have verified that the inverse situation is also true. The non-graded
versions of the graded $19$-vertex models are in correspondence with the $19$%
-vertex models of Izergin-Korepin and Zamolodchikov-Fateev.

During the preparation of this paper we learned that the connection between
Izergin-Korepin and $osp(1|2)$ models has recently been discussed in Saleur
and Wehefritz-Kaufmann \cite{Saleur}, where also earlier references are
given.

\section{Conclusion}

In the first part of this paper we have applied the coordinate {\small BA}
to find the spectra of open spin-1 chain Hamiltonians associated with four $%
19$-vertex models, including two graded models. This procedure was carried
out for boundaries derived from diagonal solutions of the reflection
equations.

We believe that the method here presented could also be applied for
Hamiltonians associated with higher states vertex-models. For instance, in
the quantum spin chain $s=3/2$ {\small XXZ} model we have four states: $%
\left| k[3/2]\right\rangle $, $\left| k[1/2]\right\rangle $, $\left|
k[-1/2]\right\rangle $ and $\left| k[-3/2]\right\rangle .$ It means that the
state $\left| k(1/2)\right\rangle $ can be parametrized by plane wave and
the states $\left| k[-1/2]\right\rangle $ and $\left| k[-3/2]\right\rangle $
as two and three states $\left| k[1/2]\right\rangle $ at the same site ,
respectively, multiplied by some weight functions.

These weight functions are responsible by the factorized form of the
two-body phase shifts of the {\small IK} model (\ref{ik.13}) and the $%
osp(1|2)$ model (\ref{osp.13}). In the {\small ZF} model, as well as in the $%
sl(2|1)$ model, we do not have a factored form for the two-pseudoparticle
phase shift because their weight functions (\ref{cba.32}) are constant. It
means that the state $\left| k[-]\right\rangle $ behaves exactly as two
states $\left| k[0]\right\rangle $ at the same site. This is in agreement
with the fact that the {\small ZF} model can be constructed by a fusion
procedure of two six-vertex models.

There are several issues left for future works. A natural extension of this
work is to consider the algebraic version for the {\small BA} \cite{FAN}.
Independently, it is interesting to analyse the Bethe Ansatz equations to
derive ground state properties, low-lying excitations and the thermodinamic
limit.

\vspace{0.5cm}{}

{\bf Acknowledgment:} This work was supported in part by Funda\c{c}\~{a}o de
Amparo \`{a} Pesquisa do Estado de S\~{a}o Paulo-{\small FAPESP}-Brasil,
Conselho Nacional de Desenvolvimento-{\small CNPq}-Brasil and by Coordena%
\c{c}\~{a}o de Aperfei\c{c}oamento de Pessoal de N\'{\i}vel
Superior--CAPES-Brasil.

\end{document}